\pdfoutput=1
\documentclass[aps,prl,nofootinbib,twocolumn,showpacs,superscriptaddress,letterpaper,amsmath,amssymb]{revtex4}

\usepackage{graphicx}  
\usepackage{dcolumn}   
\usepackage{bm}        
\usepackage{amssymb}   
\usepackage{feynmf}
\usepackage{slashed}
\usepackage{multirow}
\def\gsim{\lower0.5ex\hbox{$\:\buildrel >\over\sim\:$}}
\def\lsim{\lower0.5ex\hbox{$\:\buildrel <\over\sim\:$}}

\def\ie{{\it i.e.}}

\newcommand{\be}{\begin{equation}}
\newcommand{\ee}{\end{equation}}
\newcommand{\bea}{\begin{eqnarray}}
\newcommand{\eea}{\end{eqnarray}}

\newcommand{\nbox}{{\,\lower0.9pt\vbox{\hrule \hbox{\vrule height 0.2 cm
\hskip 0.2 cm \vrule height 0.2 cm}\hrule}\,}}

\def\missET {{\not\!\! E_T}}

\usepackage{accents}
\newlength{\dhatheight}

\def\bea{\begin{eqnarray}}
\def\eea{\end{eqnarray}}
\def\ltap{\ \raise.3ex\hbox{$<$\kern-.75em\lower1ex\hbox{$\sim$}}\ }
\def\gtap{\ \raise.3ex\hbox{$>$\kern-.75em\lower1ex\hbox{$\sim$}}\ }

\usepackage{color}

\begin{document}

\thispagestyle{empty}
\vspace*{-3.5cm}

\vspace{0.5in}

\begin{flushright}
FERMILAB-PUB-13-529-T
\end{flushright}
\title{Systematically Searching for New Resonances \\at the Energy Frontier using
   Topological Models}

\begin{center}
\begin{abstract}
We propose a new strategy to systematically search for new physics
processes in particle collisions at the energy frontier.  An examination of all possible topologies which
give identifiable resonant features in a specific
final state leads to a tractable number of `topological models' per final state and
gives specific guidance for their discovery.  Using one specific final state, $\ell\ell jj$, as an
example, we find that the number of possibilities is reasonable and
reveals simple, but as-yet-unexplored, topologies which contain
significant discovery potential.  We propose analysis techniques and
estimate the sensitivity for $pp$ collisions with $\sqrt{s}=14$ TeV and
$\mathcal{L}=300$ fb$^{-1}$.
\end{abstract}
\end{center}

\author{Mohammad Abdullah}
\author{Eric Albin}
\author{Anthony DiFranzo}
\author{Meghan Frate}
\author{Craig Pitcher}
\author{Chase Shimmin}
\author{Suneet Upadhyay}
\author{James Walker}
\author{Pierce Weatherly}
\affiliation{Department of Physics and Astronomy, University of California, Irvine, CA 92697}
\author{Patrick J. Fox}
\affiliation{Fermi National Accelerator Laboratory, Batavia, IL 60615}
\author{Daniel Whiteson}
\affiliation{Department of Physics and Astronomy, University of
  California, Irvine, CA 92697}

\maketitle


Collisions of particles at the energy frontier offer enormous potential for the discovery
of new particles or interactions.   To date, no evidence for physics beyond
the standard model has been reported.  However, the current  program consists
overwhelmingly of searches for specific theoretical
models, and the possibility remains of a theoretically
unanticipated discovery.

Searches for new particles without the guidance of a specific theory
model face daunting challenges, the most prominent being the enormous
space of signatures in which to search. An examination of every
possible observable in every final state configuration suffers from an
enormous trials factor, such that discovery is nearly impossible
unless the signal is enormous.  Previous approaches~\cite{cmsmusic,cdfsleuth,h1search} have been to consider
only the total yields in a large set of final states -- which significantly reduces the discovery sensitivity, or to search for
excesses in high-$p_T$ tails of many distributions in many final
states -- which suffer from poor statistics and large systematic uncertainties.

In this paper, we propose a new approach which focusses on exploring the complete set of models which have identifiable resonant features: for each final state, systematically construct
all possible topologies which would give resonant features, seen as
peaks in reconstructed invariant mass distributions\footnote{There are some cases in which resonances may not be reconstructable, such as compressed mass spectra where the decay products are too soft to be observed, or resonances with invisible decay products.}. This reasonable experimental
requirement constrains
the space of discoverable models dramatically without
being dominated by theoretical prejudice, and guides the analysis
strategy: to search for the resonant features of each
specifically constructed topological model. This does not completely evade the trials factor, which is unavoidable if one examines many final states, but is an experimentally motivated strategy for reducing the number of observables in a given final state.  Limits or discoveries may
be reported in terms of cross sections in the space of particle masses.
 Perhaps most importantly, the  topological model  strategy emphasizes 
{\it completeness}, where theoretical motivations may not have
inspired us to examine specific topologies.   For example, we describe below how topological models motivate a search for $Z'\rightarrow \chi_1\chi_2\rightarrow \ell^+\ell^- jj$ which features resonances in $m_{\ell\ell}$, $m_{jj}$ and $m_{\ell\ell jj}$; this is similar to existing searches for new resonances decaying to $WZ\rightarrow \ell^+\ell^- jj$ but without the constraint that $m_{\ell\ell} \approx m_Z$ and $m_{jj}\approx m_W$.  In this way, it points to
new directions where discovery potential is untapped in the current data.

This approach shares a motivational principle with the simplified model approach~\cite{Alves:2011wf},
 in that it seeks to
characterize our knowledge in terms of particle masses
 rather than
theoretical parameters, giving limits on cross sections for given
decay modes rather than on theoretical parameters for full
theories. However, while simplified models reduce the complexity of a set of full models by
specifying the minimal particle content of a topology, the  
 topological model strategy aims to cover the complete experimental space of a
particular final state or set of final states. 

This  topological model approach is best considered as an extension of the effective field
theory strategy, where the new particles are considered to be too
heavy to be directly observed as resonances and the interactions they
mediate are replaced by effective operators which integrate out the
details of the complete theory.  Typically, exhaustive lists of possible
operators are formed, giving a completeness to the analysis. In the
same way, the  topological model approach we describe here seeks to compile the list
of potentially discoverable new physics topologies, but where the new particles are light
enough to be directly seen as resonances.

The completeness of a survey of topological models in a final state gives more weight to a negative result: if nothing is seen in the data we can say with some confidence that no new observable resonances exist. In addition, another strength is it helps to organize the experimental results, which are currently presented in the context of searches for specific theories.

The number of topologies grows quickly with the number of final state
objects. For concreteness, we choose a final state with a reasonable
but non-trivial number of objects: $\ell^+\ell^- jj$.   Note that in the examples below we have focused on $\ell=e,\mu$, quark-originated jets, and that in the event selection we allow more than two jets in order to improve the probability to locate the two jets of interest among those generated by radiation.  Additionally, one could consider further categorization by jet flavor.

Though we
examine a single final state here, one important advantage of this
approach is that it allows the coherent consideration of multiple
final states. For example, one could consider $4\ell$ or $4j$ modes of $Z'\rightarrow\chi_1\chi_2$ decay.
 By contrast, a similar interpretation of multiple final-state-specific
simplified-model results is not possible, as the results are reported per final-state and the necessary correlational information needed to combine final sates is not typically reported.

In the following sections, we construct the list of  topological models for $\ell^+\ell^- jj$, detail theoretical models built to describe the phenomenology of those topologies, and present LHC sensitivity studies for each topology.

\section{Topologies in $\ell^+\ell^- jj$}

\begin{figure}
\includegraphics[width=2.5in]{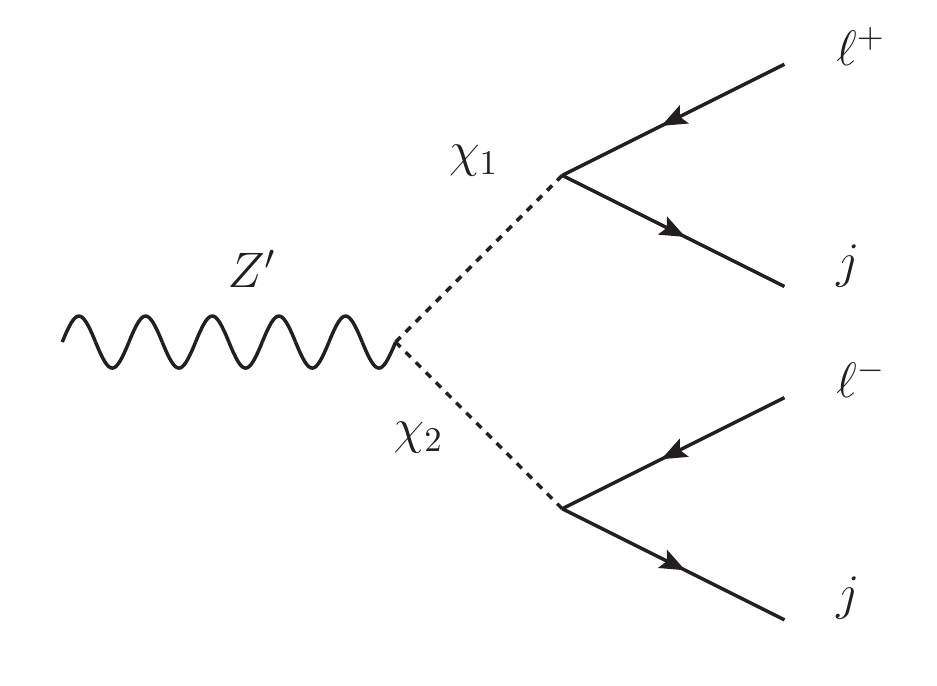}
\caption{Diagram for $Z' \rightarrow \chi_1\chi_2 \rightarrow \ell^+j\ \ell^-j$}
\label{diag_lj_lj}
\end{figure}

\begin{figure}
\includegraphics[width=2.5in]{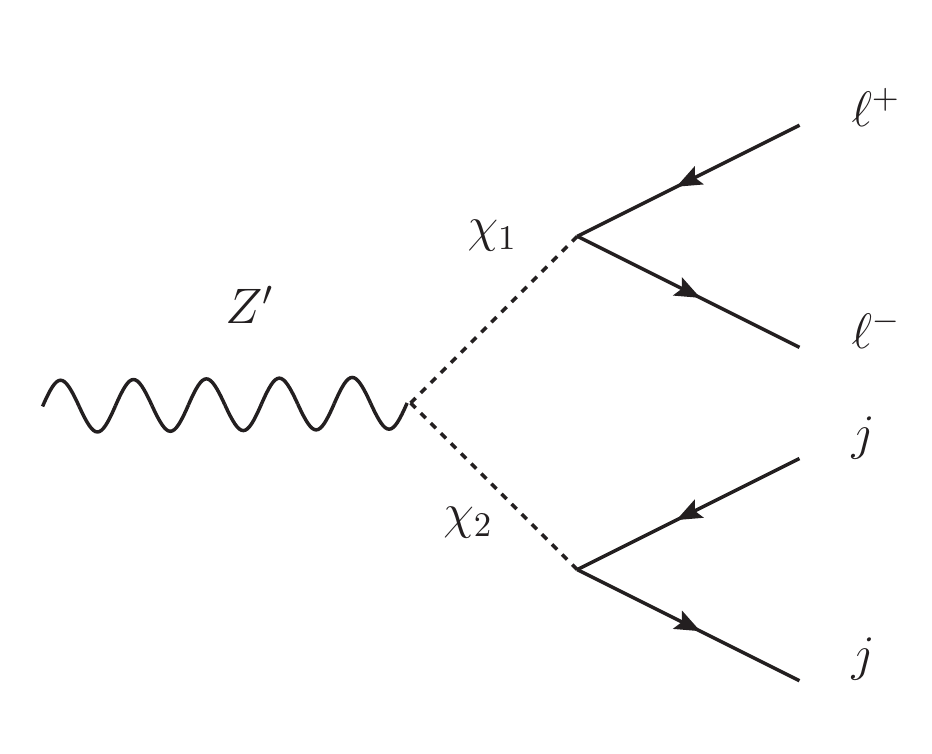}
\caption{Diagram for $Z' \rightarrow \chi_1\chi_2 \rightarrow \ell^+\ell^-\ jj$}
\label{diag_ll_jj}
\end{figure}

\begin{figure}
\includegraphics[width=3in]{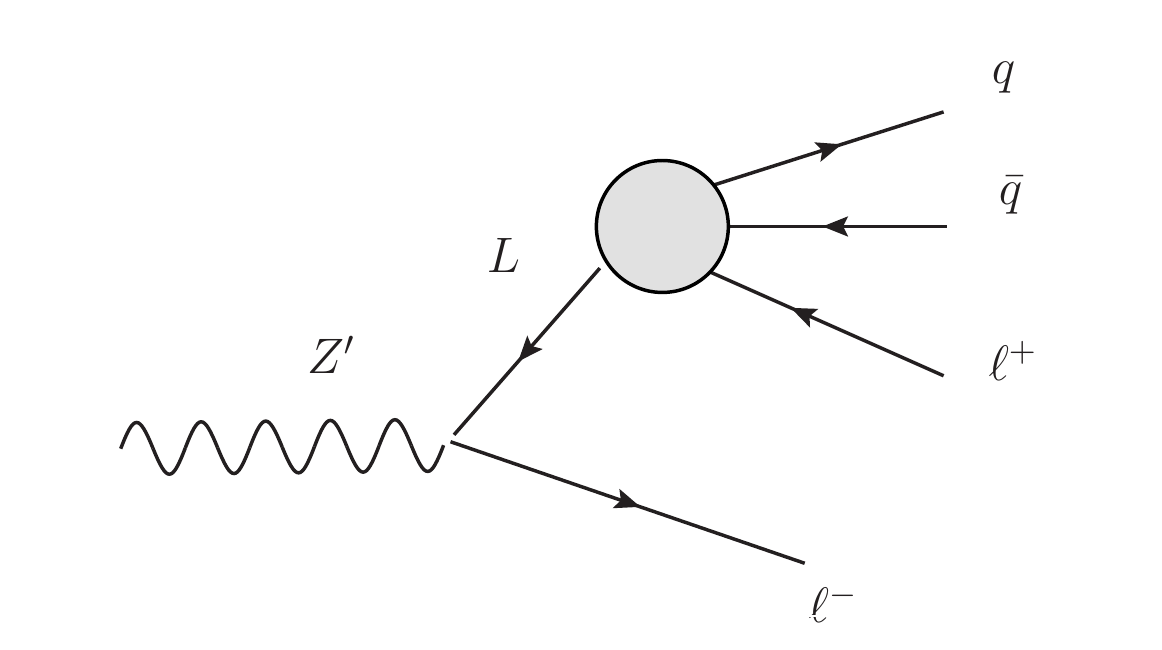}
\caption{Diagrams for $Z'\rightarrow \ell^{\mp}L \rightarrow \ell^{\mp}\ (\ell^{\pm}jj)$ (right)}
\label{diag_ljj_l}
\end{figure}

\begin{figure}
\includegraphics[width=3in]{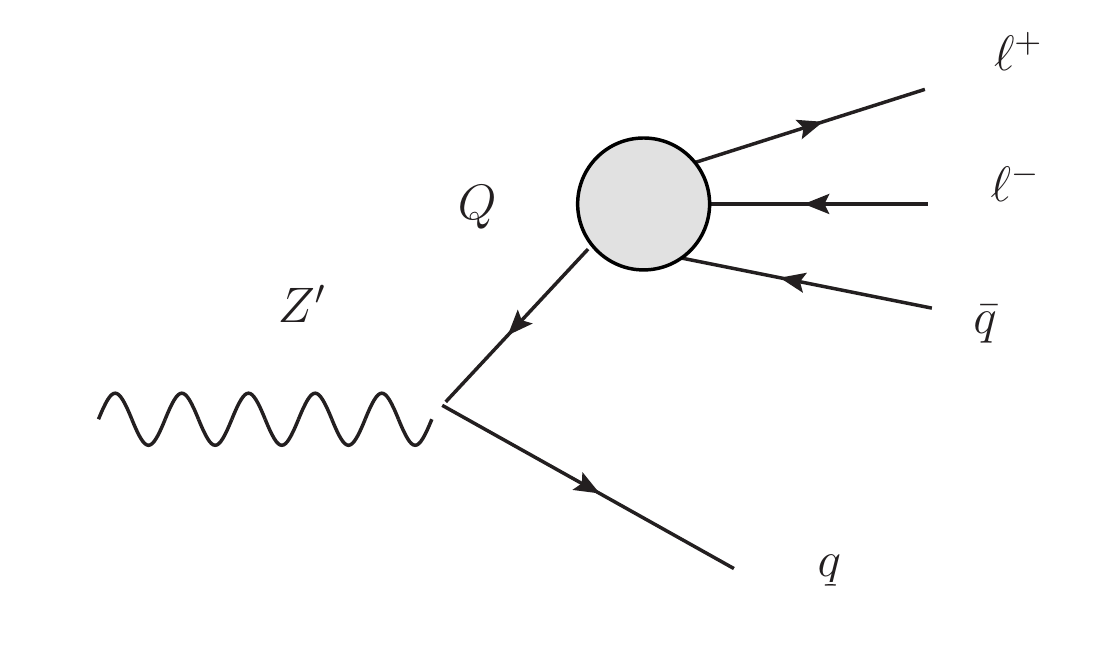}
\caption{Diagram for $Z'\rightarrow jQ \rightarrow j\ (\ell^+\ell^-j)$}
\label{diag_llj_j}
\end{figure}

In the $\ell^+\ell^- jj$ final state, resonant features can be seen in
several distinct topological arrangements; there can be two-, three-,
or four-body resonances present.  Since it will allow for the largest
production cross section, we always consider the case of a 4-body
$\ell^+\ell^- jj$ resonance.  From a $q\bar{q}$ initial state this
first resonance must be either a color singlet or octet, Lorentz
vector or scalar, and we focus here on the simple case of
color-singlet vector boson, a $Z'$ boson.  In this case the intermediate 2- (3-) body resonances are scalars (fermions).  In particular, the topologies which arise are:

\begin{itemize}
\item $Z'\rightarrow \chi_1\chi_2\rightarrow (\ell^+ j)(\ell^- j)$ see Fig.~\ref{diag_lj_lj}
\item $Z'\rightarrow \chi_1\chi_2\rightarrow (\ell^+\ell^-)(jj)$ see Fig.~\ref{diag_ll_jj}
\item $Z'\rightarrow \ell^{\pm}L\rightarrow \ell^{\pm}(\ell^{\mp}jj)$
  see Fig.~\ref{diag_ljj_l}
\item $Z'\rightarrow jQ\rightarrow j(\ell^+\ell^-j)$ see Fig.~\ref{diag_llj_j}
\end{itemize}

The first  topology (see Fig.~\ref{diag_lj_lj}) describes resonant
production of new particles, each of which
decays to a lepton and a jet. This is essentially a lepto-quark model~\cite{Buchmuller},
and territory which is well-explored experimentally and will not be discussed further
here.

The second topology (see Fig.~\ref{diag_ll_jj}) describes  production
of new particles $\chi_1$ and $\chi_2$ which decay to lepton pairs or quark pairs. This is similar
to searches for heavy resonances which decay to $ZZ$ with one hadronically
and one leptonically decaying $Z$ boson, or decays to $WZ$ with hadronic $W$-boson and leptonic $Z$-boson decays. This is well-explored territory,
but only for the cases where $\chi_1$ and $\chi_2$ have masses close to $m_{W,Z}$. In
the case when these intermediate particles are heavier, \emph{the
experimental data have not been examined, and discovery potential remains.}
This topological approach therefore provides a useful and natural
generalization of an existing effort. 

The final two topologies (see Figs.~\ref{diag_ljj_l} and~\ref{diag_llj_j}) include 
decays $Q\rightarrow \ell^+\ell^-j$ or
$L\rightarrow \ell^{\pm}jj$. These arise from a higher dimension four-fermion contact
operator, representing the mediation of this interaction via some new
heavy particle which is integrated out.  To be perfectly exhaustive, one should
consider the case where the mediator is light enough to be produced
on-shell, giving, for example $Q\rightarrow Xj
\rightarrow\ell^+\ell^-j$.

Note that we do not discuss in detail $Z'\rightarrow (\ell^+\ell^- jj)$ through
an effective five-point interaction, because this topology provides no further
intermediate resonances to guide the search beyond $m_{\ell\ell jj}$.

In the next sections, we construct example models which give these
topologies, propose techniques for experimental analysis, and
estimate the sensitivity of the LHC at $\sqrt{s}=14$ TeV, with $\mathcal{L}=300$ fb$^{-1}$.

\section{The Models}

In order to allow simulation of these topologies, toy models were built in FeynRules \cite{Christensen:2008py} for MadGraph~\cite{madgraph}.  To allow production, the $Z'$ resonance must couple to some of the SM quarks.  This may occur either through charging some of the quarks under the $U(1)'$, or through a higher dimension operator as an ``effective $Z'$" \cite{Fox:2011qd}.  The former requires the addition of new heavy fermions to cancel gauge anomalies, these can always be vector-like under the SM and chiral under the $U(1)'$ \cite{Batra:2005rh},  The latter requires heavy fermions that mix with SM fermions, these can be vector-like under both the SM and the $U(1)'$.  We consider a flavor independent coupling of the RH quarks to the $Z'$.  Since it does not affect the physics we are interested in, our toy models are not complete and do not contain either sets of heavy fermions necessary to make the models consistent.  Furthermore, we assume the Higgs field that is responsible for breaking the $U(1)'$ is sufficiently massive that it does not play a role in the following.
How the $Z'$ decays  determines the final state topology; each case is discussed below.

\subsection{$(\bar{q}q)(\ell \ell)$ Topological Model}

This toy model contains two new scalars $\phi_1$ and $\phi_2$, that are charged
under the $Z'$. These $\phi$ fields are not the
mass eigenstates, instead they mix with one another to give the mass eigenstates $\chi_1$ and $\chi_2$.  In general there will be decays of $Z'$ to all open channels \ie $\chi_1\chi_1$, $\chi_1\chi_2$, and $\chi_2\chi_2$.  However, by judicious choice of the scalar charges and mixing angles, it is possible to build a toy model in which the $Z'$ decays are restricted to $jj$ along with $\chi_1\chi_2$ only, or $\chi_1\chi_2$ and one of $\chi_1 \chi_1$ or $\chi_2\chi_2$.

To allow the scalars to decay we turn on some higher dimension operators of the form $\lambda^k_{ij}\frac{\phi_k X^n}{\Lambda^{1+n}}\ell_i H e^c_j$ where $n$ is chosen to make the operator $U(1)'$ invariant and $\langle X\rangle \ne 0$.  The $\lambda$ coefficients are a new source of flavor changing processes mediated by the $\phi$'s, for which there are strong constraints.  If the $\lambda$ are taken proportional to the SM Yukawa matrices then the couplings of $\phi$ will be diagonal in the quark/lepton mass basis, which then means the $\chi$'s preferentially decay to heavy flavor.  However, since we do not take advantage of flavor tagging in this analysis, we make a simplifying assumption that instead the couplings of $\chi$ are flavor universal in the quark and lepton mass bases.

\begin{figure}
\includegraphics[width=0.75\linewidth]{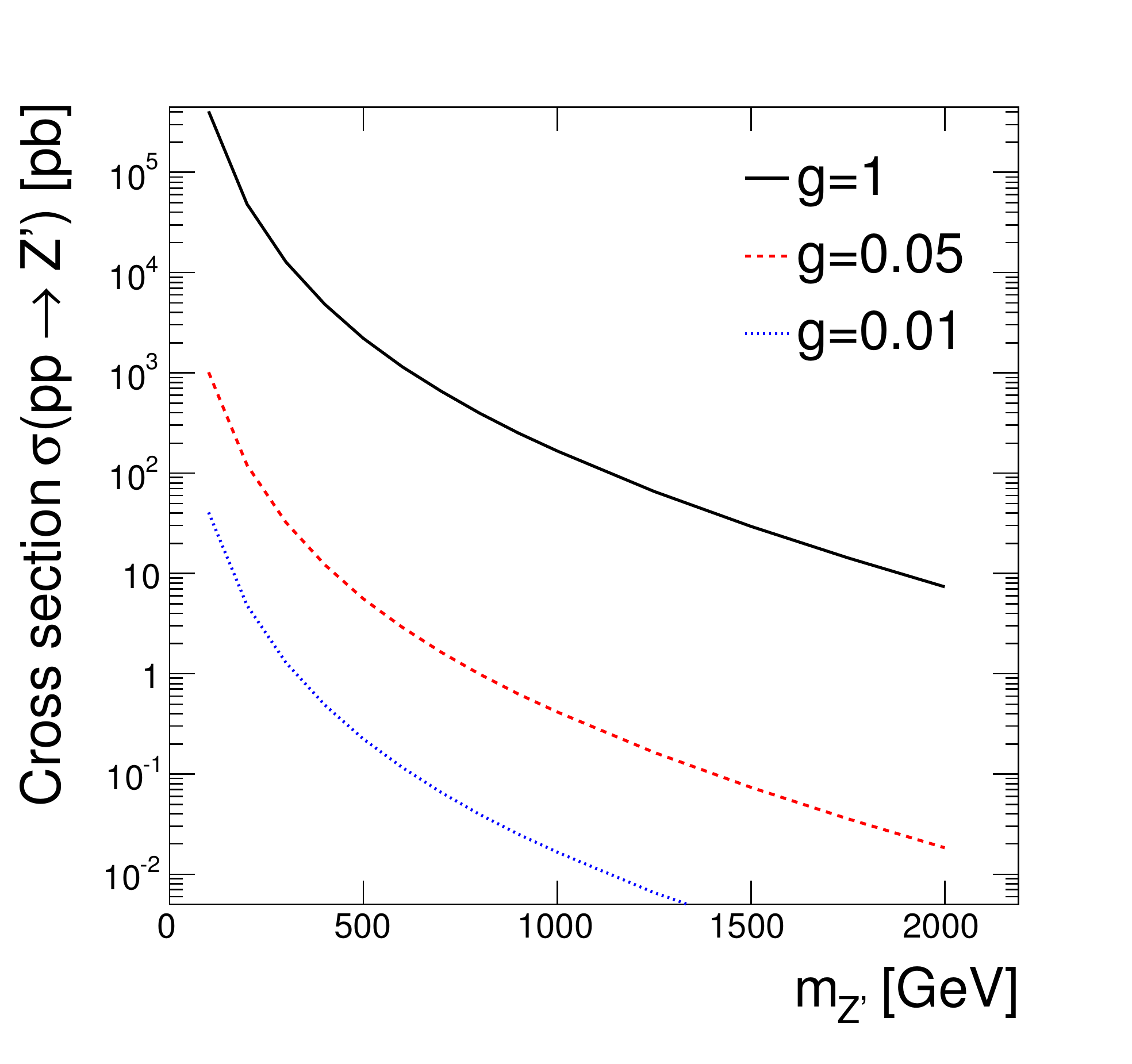}
\caption{Cross section for $Z'$ production at $\sqrt{s}=14$ TeV, 
  including all decay modes.}
\label{fig:xsec}
\end{figure}

Cross section for $Z'$ production in $pp$ collisions at \mbox{$\sqrt{s}=14$}
TeV is given in Fig~\ref{fig:xsec}.

\subsection{$(\bar{q}q\ell)\ell$ Topological Model}

A model to describe this resonant structure begins with a new
$Z'$ boson, as above. In addition, we  include heavy
vector-like pairs of leptons $(L,\,L^c)$ that will mix with the right-handed
standard model leptons.  $L^c$ has the quantum numbers of the RH leptons under the SM, \ie\ $Y=1$, and is charge 1 under the $U(1)'$, the $L$ is the opposite under both.  This mixing is induced through operators involving the field, $X$, responsible for the $Z'$ mass, $\lambda X L_i e^c_i$. To avoid lepton flavor constraints we have included three generations of heavy vector-like leptons and they mix with the SM RH leptons in a flavor universal fashion.  However, when investigating the reach at the LHC in this final state we will focus exclusively on $\ell=e$.  This mixing term also allows the heavy leptons to decay, through an off shell $X$ field.  We introduce higher dimension couplings of $X$ to (down-type) quarks of the form $\lambda'_{ij}\frac{X}{\Lambda}q_i H d^c_j$ , so that $L\rightarrow \ell \bar{q}q$ and, as before, we take this coupling to be universal in the mass basis of the quarks.  Thus, the toy model under investigation is one in which the possible final states of the $Z'$ are dijets, $L^+ L^-$, $L^\pm e^\mp$, and $e^+e^-$, with the heavy leptons decaying as $L^\pm\rightarrow e^\pm q\bar{q}$.  For small mixing, $\epsilon$, the couplings of these last three states are in the ratio $1:\epsilon:\epsilon^2$, so the strongly constrained dilepton rate can made parametrically small whilst maintaining an interesting rate in $L^\pm e^\mp$.


\subsection{$(q\ell\ell)q$ Topological Model}

A model describing the resonant structure $(q\ell\ell)q$ begins as
well with a new $Z'$ boson and is very similar to the model for
$(q\bar{q}\ell)\ell$, but the heavy lepton $L$ is replaced by a heavy
quark, $Q$, taken to have the SM quantum numbers of the $d_R$, and $X$
decays to leptons, Fig. \ref{diag_llj_j}.  The final states of the $Z'$ are dijet, $Q\bar{q}$, and $Q\bar{Q}$ if it is kinematically accessible.  Thus, in addition to the final state we are interested in there will be constraints from dijet resonance searches as well as strong constraints from multilepton searches.  Searches for a heavy quark decaying as $Q\rightarrow q Z$ can also be recast~\cite{Cranmer:2010hk} to place bounds on this model.



\section{Backgrounds to $\ell\ell jj$}

In $pp$ collisions, the dominant background in the $\ell\ell jj$ final
state to any of these new models
is  $Z/\gamma\rightarrow \ell^+\ell^-$ in
association with jets.  Secondary backgrounds include diboson
production ($WZ \rightarrow qq'\ell^+\ell^-$ and $ZZ \rightarrow
q\bar{q}\ell^+\ell^-$) and top-quark pair production
($t\bar{t}\rightarrow W^+bW^-\bar{b}\rightarrow \ell^+\nu
b\ell^-\nu\bar{b}$). Other sources, such as $W\rightarrow \ell^\pm\nu$
in association with jets where one jet is misidentified as a lepton,
are minor in comparison and neglected here.

\begin{figure}
\includegraphics[width=3in]{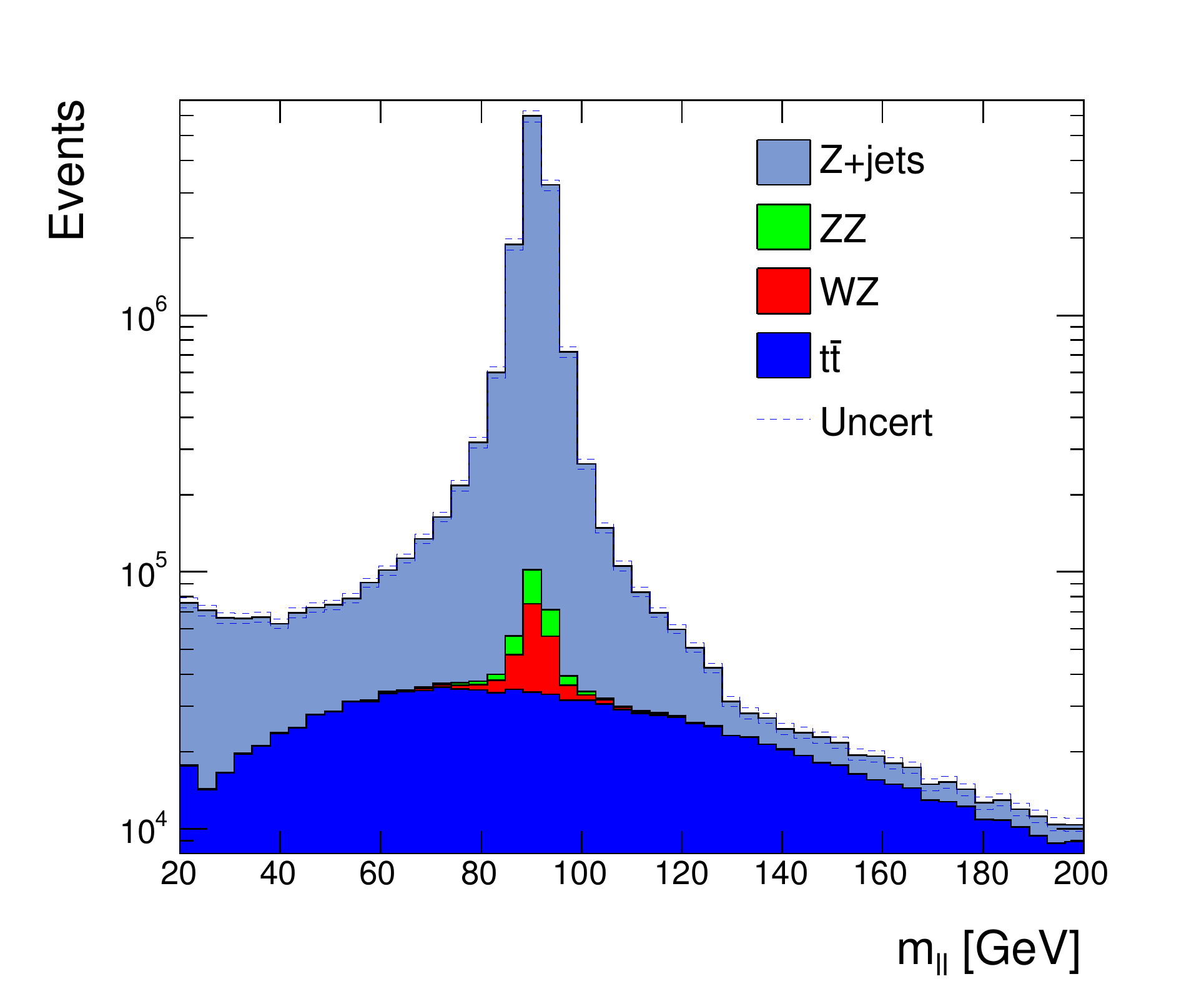}
\caption{Distribution of $m_{\ell^+\ell^-}$ in simulated events for
  background process contributing to the $\ell\ell jj$ final state in
  $pp$ collisions at $\sqrt{s}=14$~TeV with $\mathcal{L}=300$~fb$^{-1}$,
  after preselection requirements.}
\label{fig:presel}
\end{figure}

Events are simulated with {\sc madgraph5}~\cite{madgraph} with {\sc
  pythia}~\cite{pythia} for showering and hadronization and {\sc
  delphes}~\cite{delphes} for detector simulation. Table~\ref{tab:presel}  and Fig.~\ref{fig:presel}
shows the expected background yields at this stage.  Next-to-leading
order cross sections are used in each
case~\cite{Campbell:2011bn,Binoth:2010ra,Aliev:2010zk}.  Throughout, limits are
 calculated using a frequentist asymptotic calculation~\cite{roostats,asymptotics}.

We select events which satisfy the basic event topology:

\begin{itemize}
\item exactly two electrons or two muons, both with $p_{\textrm T}>20$ GeV and
  $|\eta|<2.5$
\item at least two jets, each with $p_{\textrm T}>25$ GeV and $|\eta|<2.5$
\end{itemize}

\noindent
In addition, we require $\missET<100$ GeV to partially suppress the $t\bar{t}$
background at little cost to the signal efficiency.  For each signal
hypothesis, we make further requirements on the reconstructed
invariant masses. This basic selection we refer to as our {\it
  preselection}.

\begin{table}
\caption{Expected yield from background processes at $\sqrt{s}=14$ TeV
  with $\mathcal{L}=300$ fb$^{-1}$ after preselection
  requirements. Uncertainties are dominated by theoretical cross section
  uncertainties.}
\label{tab:presel}
\begin{tabular}{lrrr}\hline\hline
Process& $\ell\ell+$jj& $\mu\mu$+jj& $ee$+jj\\\hline
$Z$+jets & $(2.35 \pm 0.1)\cdot 10^{7}$ & $(1.3 \pm 0.1)\cdot 10^{7}$ & $(1.0 \pm 0.1)\cdot 10^{7}$\\
$ZZ$ & $(5.77 \pm 0.3)\cdot 10^{4}$ & $(3.1 \pm 0.2)\cdot 10^{4}$ & $(2.6 \pm 0.1)\cdot 10^{4}$\\
$WZ$ & $(8.68 \pm 0.4)\cdot 10^{4}$ & $(4.6 \pm 0.2)\cdot 10^{4}$ & $(4.0 \pm 0.2)\cdot 10^{4}$\\
$t\bar{t}$ & $(1.24 \pm 0.1)\cdot 10^{5}$ & $(6.6 \pm 0.4)\cdot 10^{4}$ & $(5.9 \pm 0.3)\cdot 10^{4}$\\
\hline
 Total & $(2.37 \pm 0.1)\cdot 10^{7}$& $(1.3 \pm 0.1)\cdot 10^{7}$&
 $(1.0 \pm 0.1)\cdot 10^{7}$\\\hline\hline\hline
\end{tabular}
\end{table}

\section{Mass reconstruction and LHC Sensitivity}

In each topology,  there are combinatorial ambiguities in the
assignment of reconstructed jets to colored partons~\cite{jetcomb}. In the heavy lepton model,
there is an additional ambiguity regarding which charged lepton is assigned to
the decay of the heavy lepton, see Fig~\ref{diag_ljj_l}. To resolve
ambiguities, we use the separation $\Delta
R=\sqrt{\Delta\phi^2+\Delta\eta^2}$ to either select decay products
with the smallest or largest opening angle depending on the kinematic configuration. Details are given below in each case.

\subsection{$(\ell\ell)(jj)$ topology}

\begin{figure}
\includegraphics[width=2in]{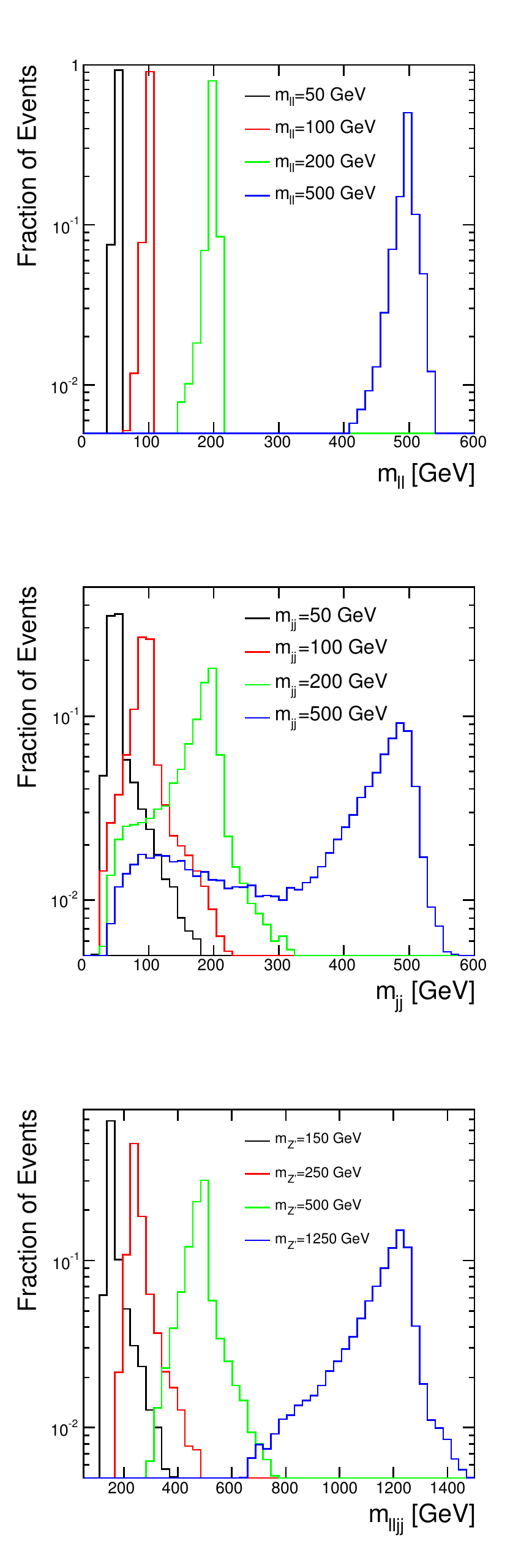}
\caption{In simulated $Z'\rightarrow \chi_1\chi_2\rightarrow \ell\ell
  jj$  events, the distribution of reconstructed invariant $\ell\ell$, $jj$ and
  $\ell\ell jj$ masses for several values of  $m_{Z'}, m_{\chi_1}$
  and $m_{\chi_2}$.  The normalization is arbitrary. The shoulder in $m_{jj}$ is due to imperfect
  selection of the jet pair.}
\label{fig:mass_lljj}
\end{figure}

In the $(\ell\ell)(jj)$ topology, there are no lepton ambiguities, so
the $(\ell\ell)$ system is well-defined.  In the case of the jets,  if more than two jets are found there are several possibilities for the
$(jj)$ system.  The $(jj)$ pair momentum balances the momentum of the
$(\ell\ell)$ pair, and we choose the pair of jets with largest $\Delta
R(\ell\ell,jj)$. Examples of reconstructed masses are shown in Fig~\ref{fig:mass_lljj}.

\begin{figure}
\includegraphics[width=2in]{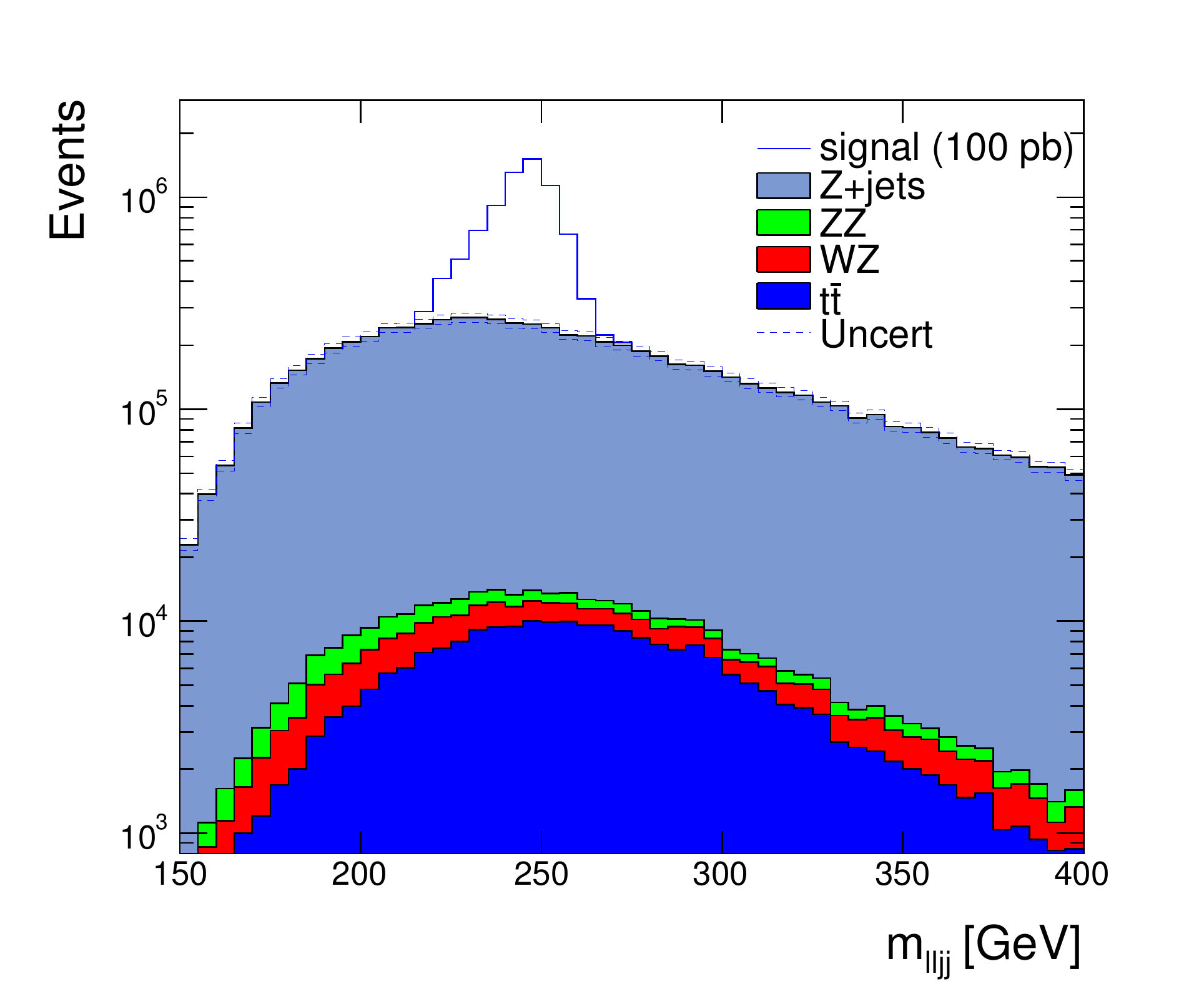}
\includegraphics[width=2in]{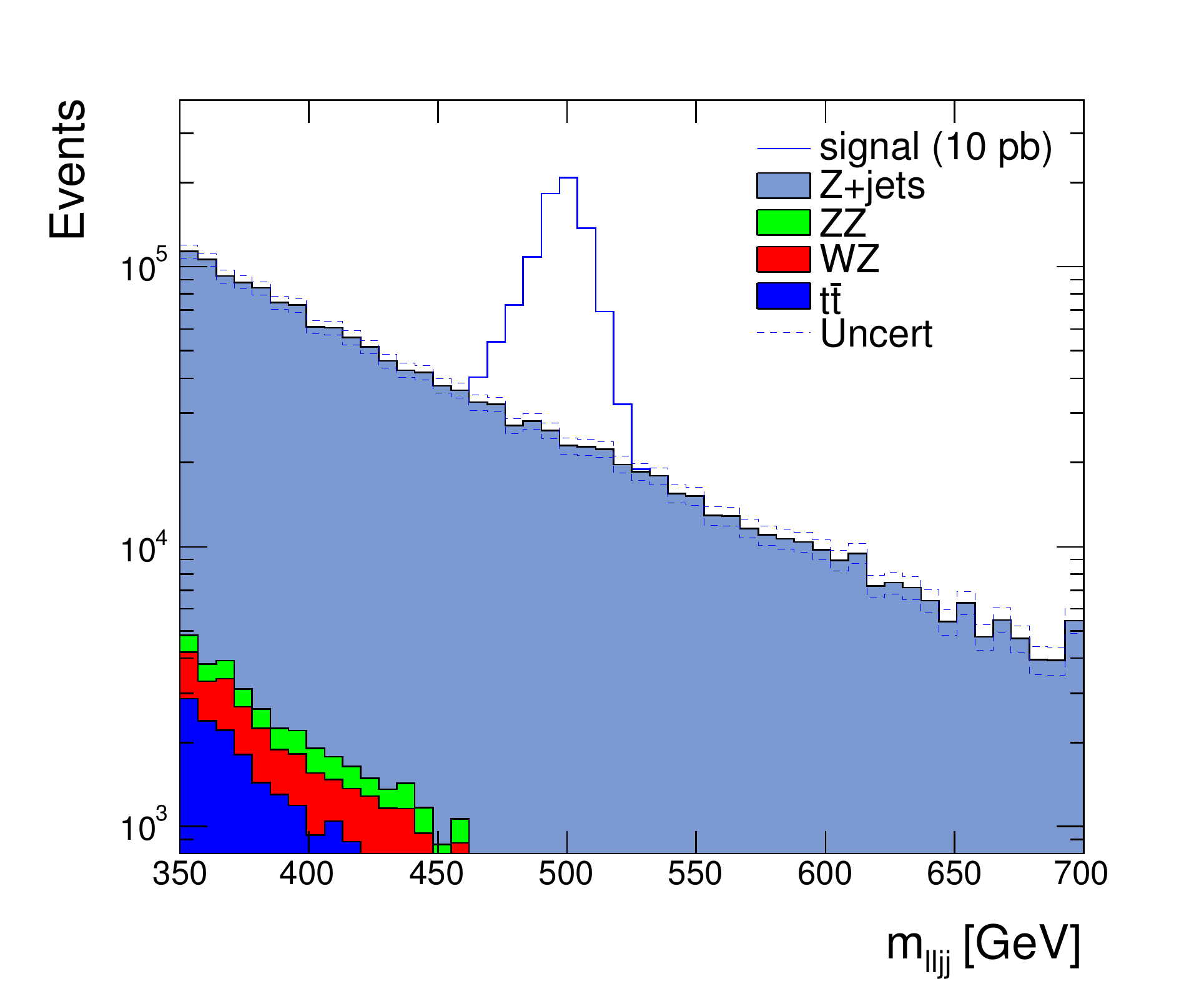}
\caption{ In the $(\ell\ell)(jj)$ topology, distribution of
  $m_{\ell\ell jj}$ in simulated signal and background
  events for two example mass points, after requirements on
  $m_{\ell\ell}$ and $m_{jj}$ in
  $pp$ collisions at $\sqrt{s}=14$~TeV with
  $\mathcal{L}=300$~fb$^{-1}$. 
Top shows the case of  $m_{Z'}=250$ GeV,
  $m_{\chi_1,\chi_2}=100$ GeV; bottom shows the case of  $m_{Z'}=500$ GeV,
  $m_{\chi_1,\chi_2}=100$ GeV. In both cases, an arbitrary value of $\sigma(pp\rightarrow Z'\rightarrow \chi_1\chi_2\rightarrow
 \ell\ell jj)$  is assumed.}
\label{fig:ex1}.
\end{figure}

\begin{figure}
\includegraphics[width=1.5in]{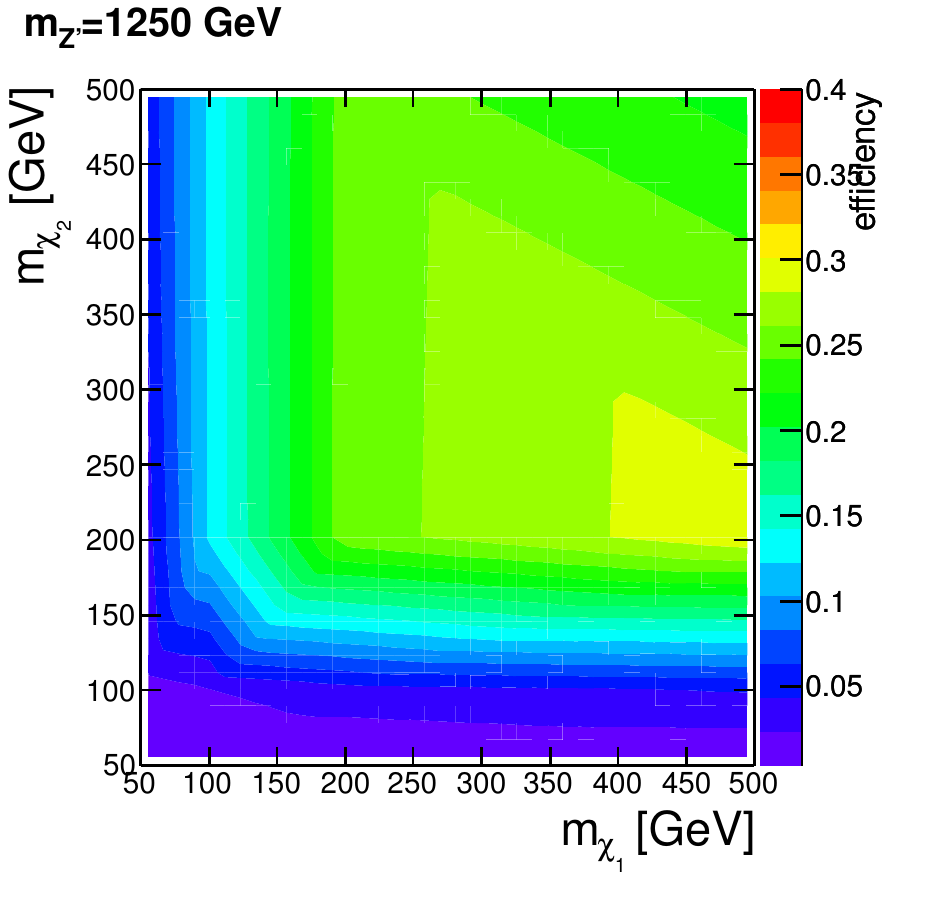}
\includegraphics[width=1.5in]{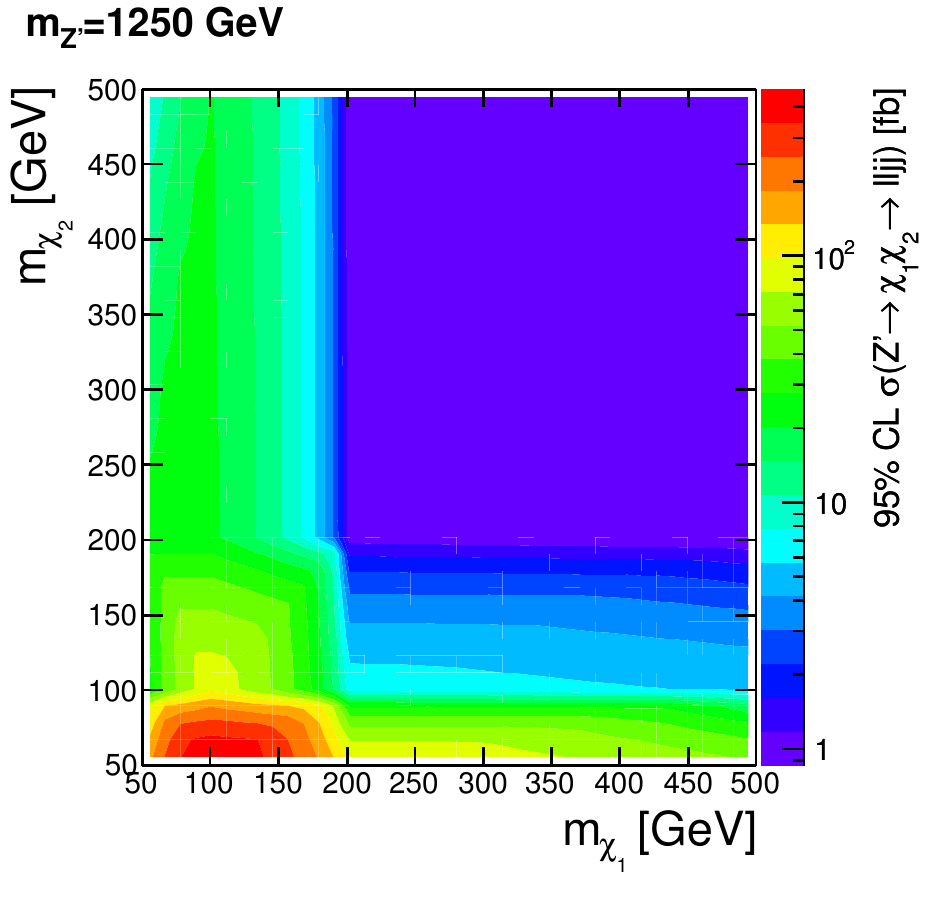}
\includegraphics[width=1.5in]{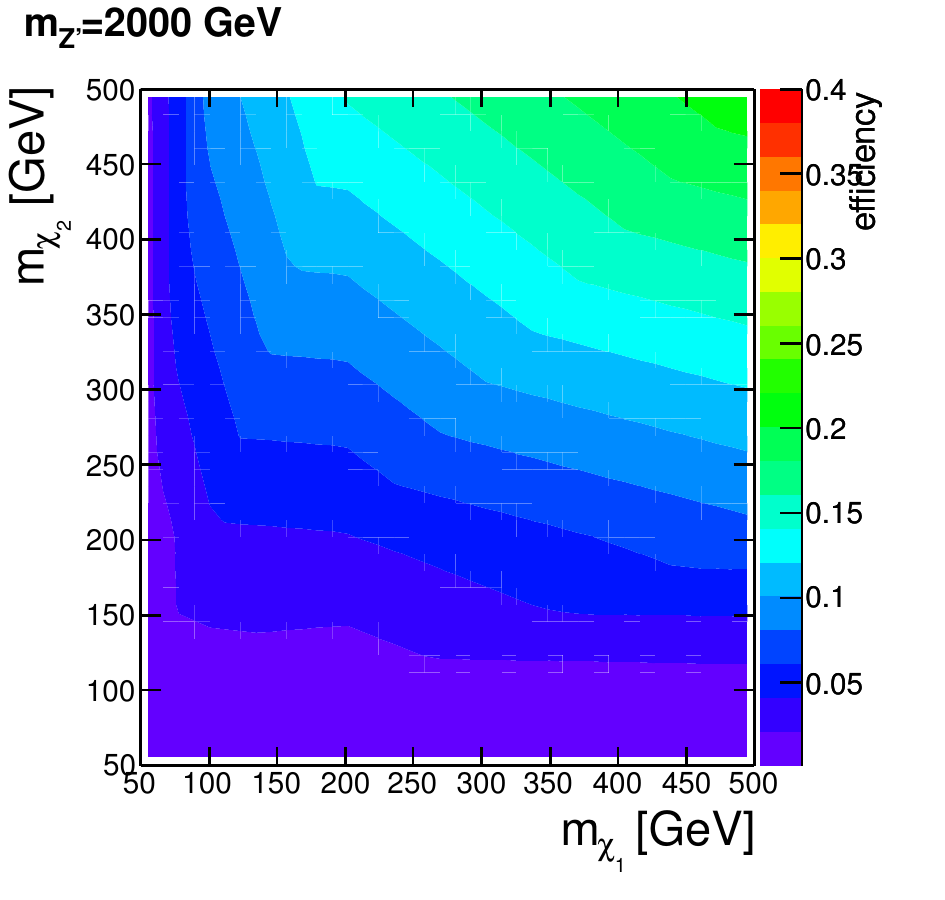}
\includegraphics[width=1.5in]{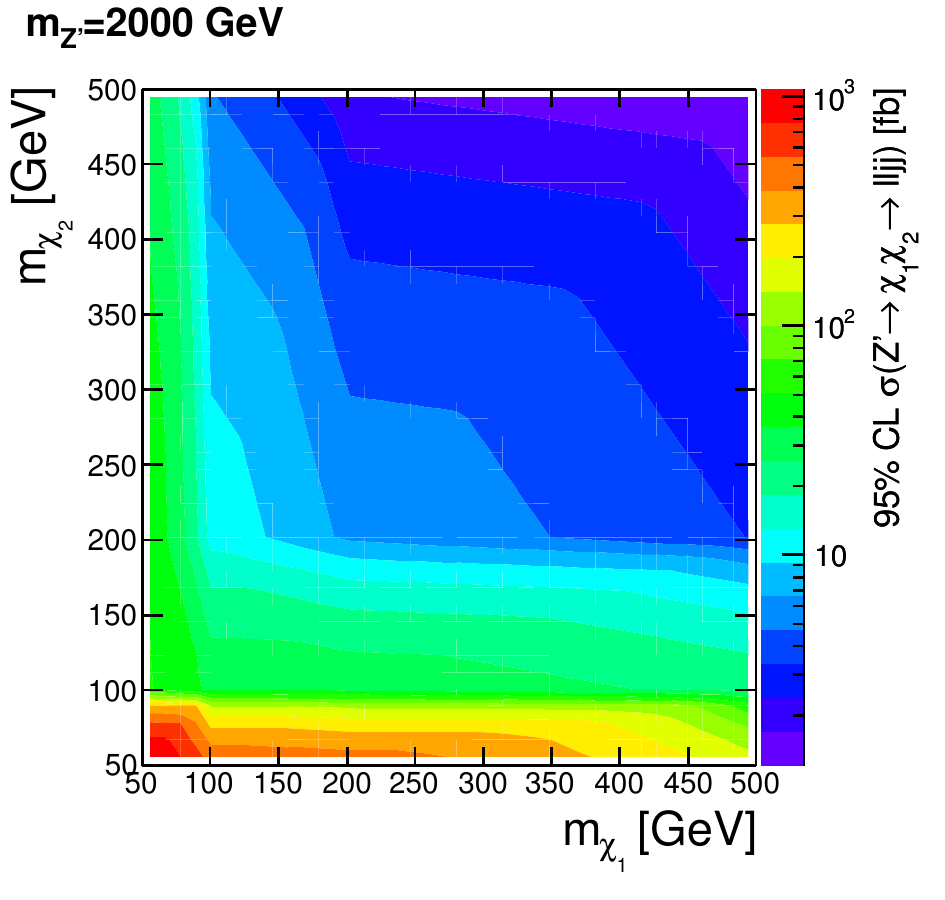}
\caption{In the $(\ell\ell)(jj)$ topology, selection efficiency and expected cross-section upper limits versus $m_{\chi_1}$ and
  $m_{\chi_2}$ for several choices of $m_{Z'}$  at $\sqrt{s}=14$ TeV
  with $\mathcal{L}=300$ fb$^{-1}$.  For small values of
  $m_{\chi_2}$, the efficiency is small due to jet $p_{\textrm T}$
  requirements and jet resolution effects. For values of $m_{\chi_1}$
  near $m_Z$, the larger backgrounds lead to weakened limits. }
\label{fig:lim_ll_jj}
\end{figure}

In addition to the preselection requirements above, we select events
with reconstructed mass values close to the true values, $m_{\ell\ell}
\in [m_{\chi_1}-25,m_{\chi_1}+25]$, $m_{jj}
\in [m_{\chi_2}-100,m_{\chi_2}+50]$, and $m_{lljj}
\in [m_{Z'}-250,m_{Z'}+100]$ GeV.  Example distributions after $m_{ll}$
and $m_{jj}$ requirements can be seen for two examples in Fig~\ref{fig:ex1}.  Efficiency of the final selection and expected upper limits on
 $\sigma(pp\rightarrow Z'\rightarrow \chi_1\chi_2\rightarrow
 \ell\ell jj)$ can be seen in Fig~\ref{fig:lim_ll_jj}.  

\begin{figure}
\includegraphics[width=2in]{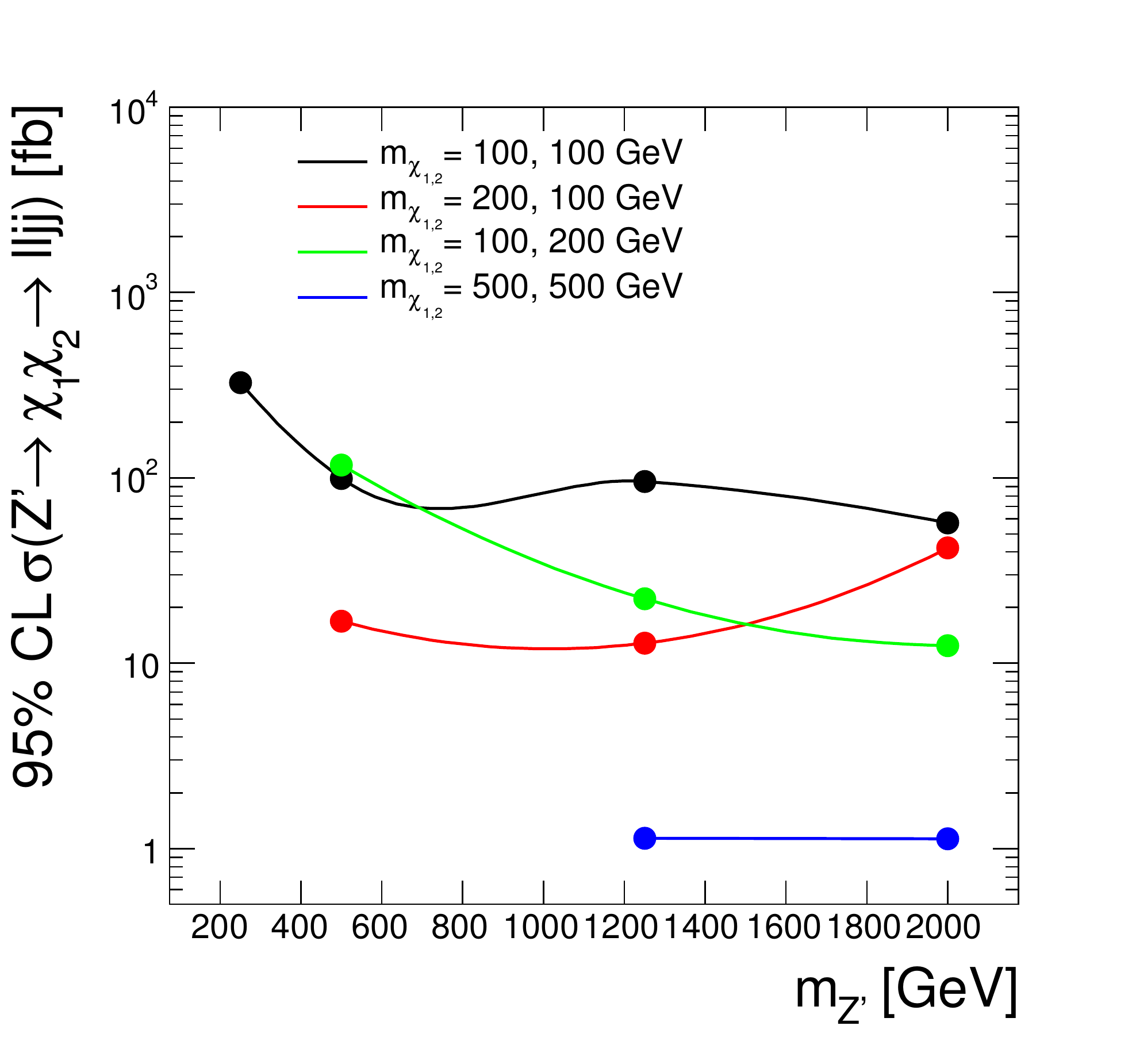}
\caption{In the $(\ell\ell)(jj)$ topology, expected upper limits on
  the cross section $\sigma(pp\rightarrow
  Z'\rightarrow\chi_1\chi_2\rightarrow \ell\ell jj)$ at 95\% CL versus $m_{Z'}$ for several
  choices of $m_{\chi_1}$ and
  $m_{\chi_2}$  in
  $pp$ collisions at $\sqrt{s}=14$~TeV with
  $\mathcal{L}=300$~fb$^{-1}$. The $m_{\chi_{1,2}}=200,100$ GeV (red)
  and $m_{\chi_{1,2}}=100,200$ curves have different dependences on
  $m_{Z'}$ due to the asymmetry in the lepton and jet efficiencies for
  large values of $p_{T}^\chi$.}
\label{fig:mass_lim_ll_jj}
\end{figure}

\begin{figure}
\includegraphics[width=2in]{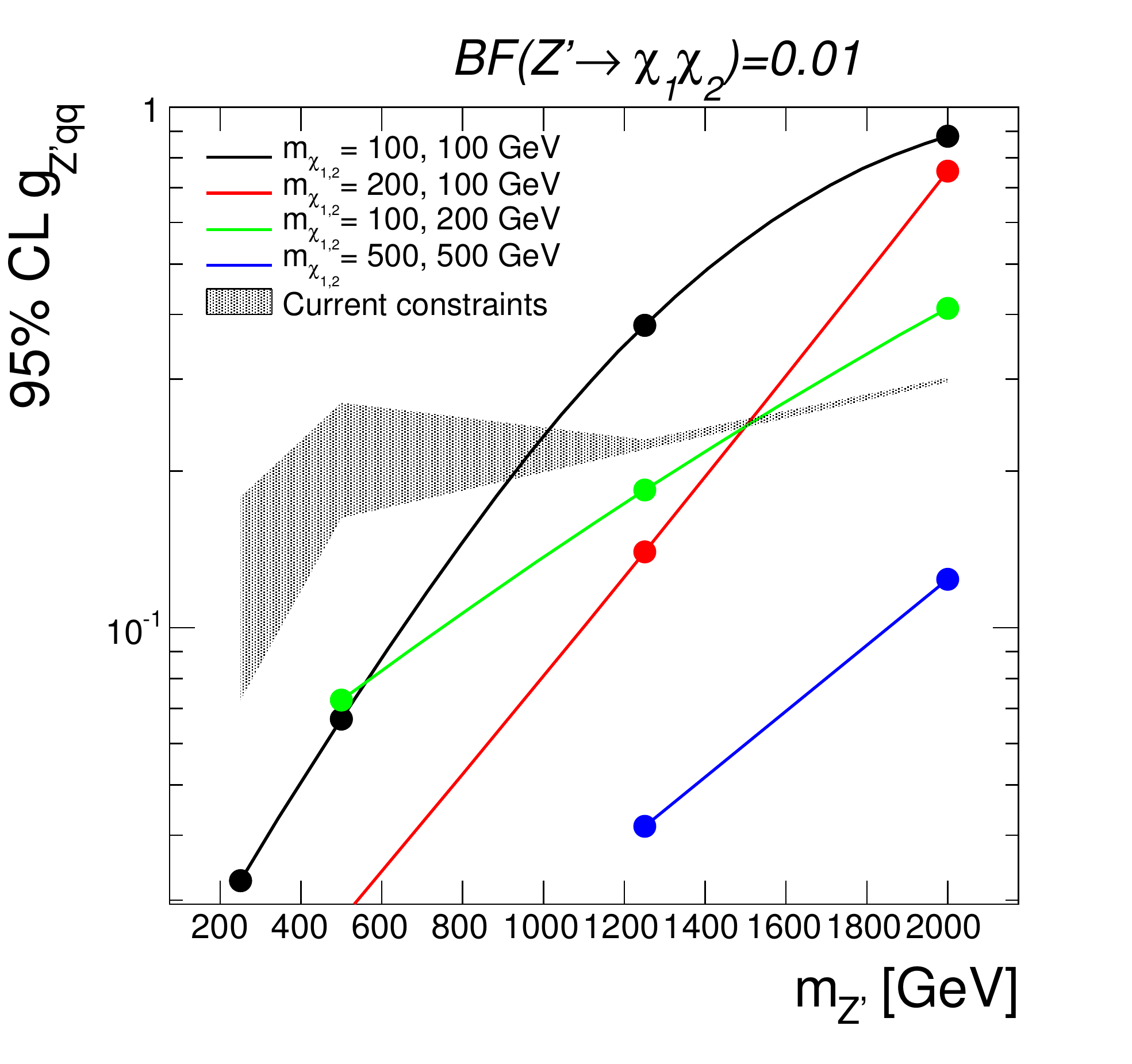}
\includegraphics[width=2in]{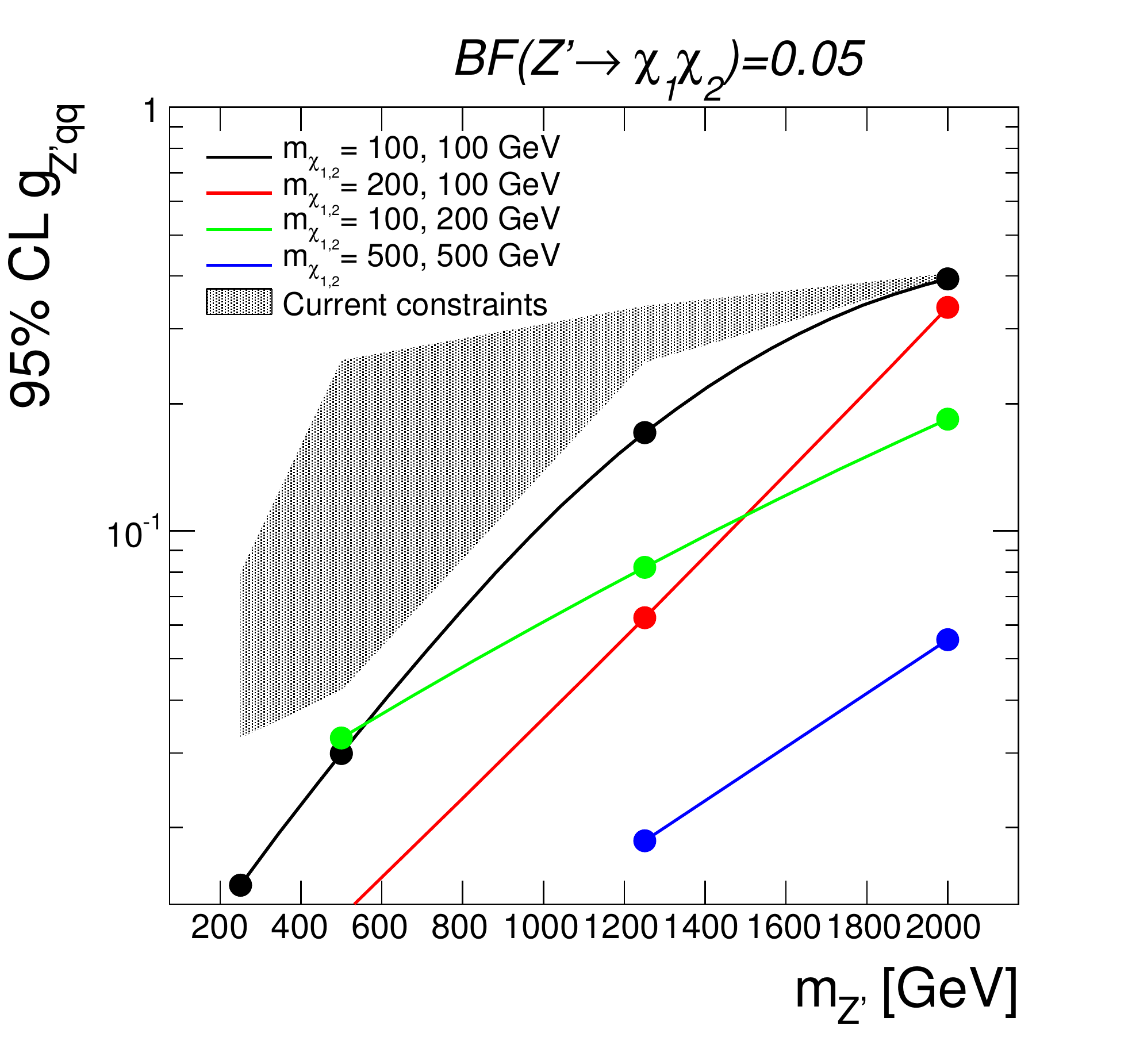}
\caption{Limits on coupling $g_{Z'qq}$ for two choices of
  BF($Z'\rightarrow\chi_1\chi_2$).   The shaded region shows the
  current limits on the coupling from other topologies (see text)
  where the width of the band reflects the variation with assumed
  $m_{\chi_1}$ and $m_{\chi_2}$ in
  $pp$ collisions at $\sqrt{s}=14$~TeV with $\mathcal{L}=300$~fb$^{-1}$.}
\label{fig:mass_limg_ll_jj}
\end{figure}

Limits  on cross section are
shown in Fig~\ref{fig:mass_lim_ll_jj} and converted into limits on the
coupling $g_{Z'qq}$ versus $m_{Z'}$  in Fig~\ref{fig:mass_limg_ll_jj}
for two choices of BF($Z'\rightarrow \chi_1\chi_2$).  The model as constructed would
give a signature in $\ell\ell jj$, but the new particles and
interactions introduced would yield signatures in other channels, where existing
limits may also constrain the parameters of this model.  Specifically,
$Z'\rightarrow \chi_1\bar{\chi_1}$ gives a $\ell\ell\ell\ell$ final
state, while $Z'\rightarrow \chi_2\bar{\chi_2}$ gives a $jjjj$ final
state and $Z'\rightarrow jj$ gives a di-jet final state.  

Both CMS \cite{cmsmultilepton} and ATLAS \cite{atlas4l} have 
SUSY-motivated searches for four leptons and with and without missing energy using the full dataset.  These results place strong constraint on our $4\ell$ final state that contains no intrinsic missing energy.  This constraint is stronger than those coming from any other channel.  However, as discussed earlier, it is possible that the mixings of the scalars are such that $Z'\rightarrow \chi_1\bar{\chi_1}$ is forbidden and this strong constraint is avoided; we focus on this possibility here.
Both the Tevatron and LHC have searched for $4j$ final states~\cite{atlas4j,cms4j,cdf4j}.  These coloron searches can be recast in terms of our model.   Finally, there are bounds on dijet resonances~\cite{cdf2j,cdf2jb,cms2j}, though exactly which analysis is strongest depends sensitively on the mass of the $Z'$ boson~\cite{Dobrescu:2013cmh}.  Using the constraints on a vector boson of a gauged baryon number, $g_B$, presented in Ref.~\cite{Dobrescu:2013cmh} our coupling is bounded by $g_{Z'qq}\ltap g_B/6\sqrt{BF(jj)}$.  

In order to compare the limits from
all these searches we must assume something about the branching ratios of the $Z'$.  As described earlier, the $Z'$ must minimally decay to $\chi_1\chi_2$ and $jj$.  Since the $4\ell$ mode is so constraining we consider the situation where this mode is forbidden at tree level and further we make the simplifying assumption that branching fractions to $\chi_1\chi_2$ and $\chi_2\chi_2$ are equal, with the remaining decay mode being back to dijets.  The constraints from the other modes, along with the results of our analysis, on the common coupling of $g_{Z'}$ are shown in Fig.~\ref{fig:mass_limg_ll_jj} for two different choices of branching fraction.

\subsection{$\ell(\ell jj)$ topology}

\begin{figure}
\includegraphics[width=2in]{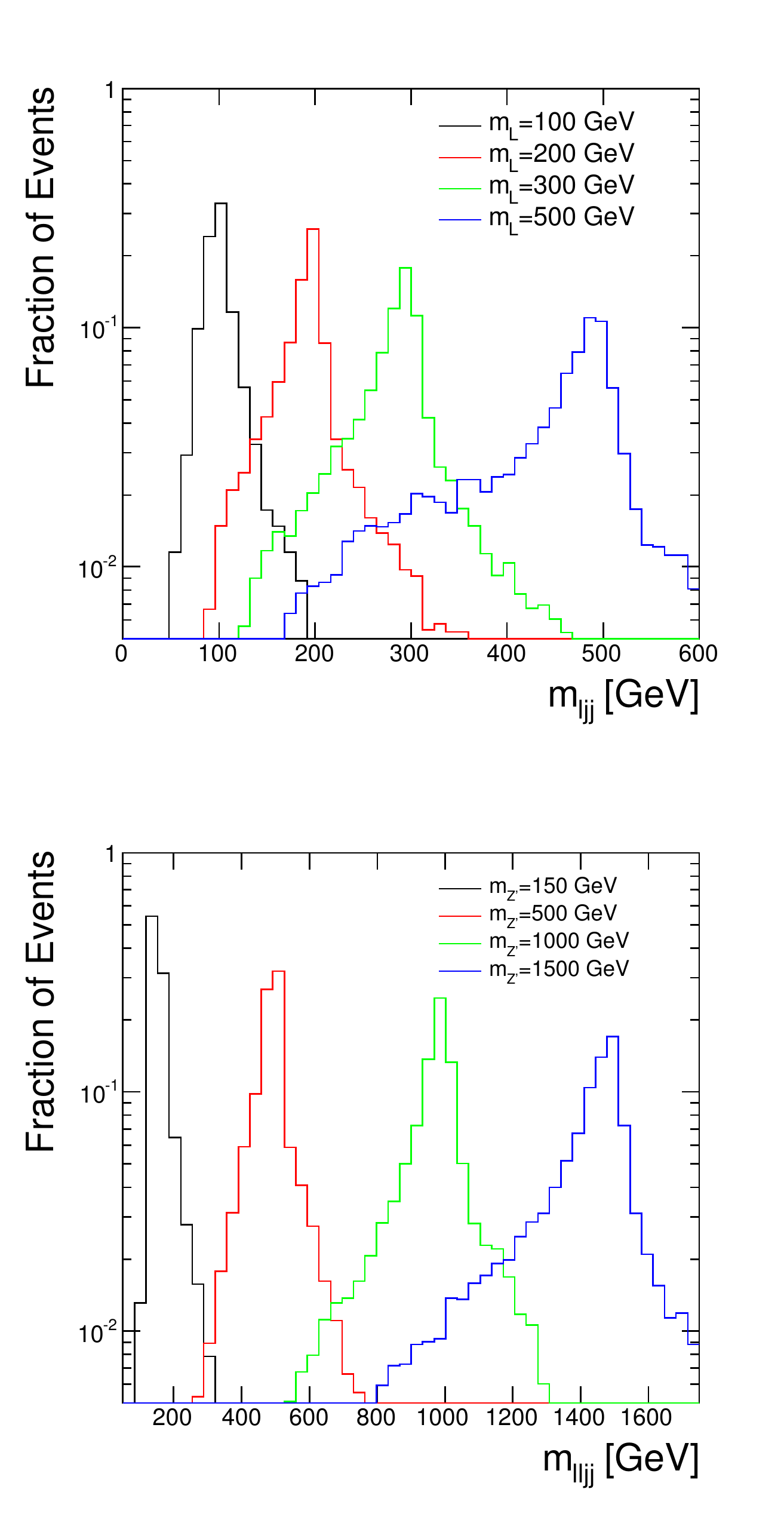}
\caption{In simulated $Z'\rightarrow \ell L\rightarrow \ell \ell jj$
  events, distribution of reconstructed invariant $\ell jj$ and
  $\ell\ell jj$ masses for several values of  $m_{Z'}$
  and $m_{L}$. Normalization is arbitrary. }
\label{fig:mass_l_ljj}
\end{figure}

Resonance reconstruction in the $\ell(\ell jj)$ topology begins with
the identification of the two jets produced in $L\rightarrow \ell jj$ decay.
 If more than two jets are found in the event, the pair with smallest $\Delta R(j,j)$
are chosen. The lepton with smaller $\Delta R(\ell,jj)$ is chosen to
form $m_{L}=m_{\ell jj}$. The second lepton then completes the
system. Examples of reconstructed masses are shown in Fig~\ref{fig:mass_l_ljj}.

\begin{figure}
\includegraphics[width=2in]{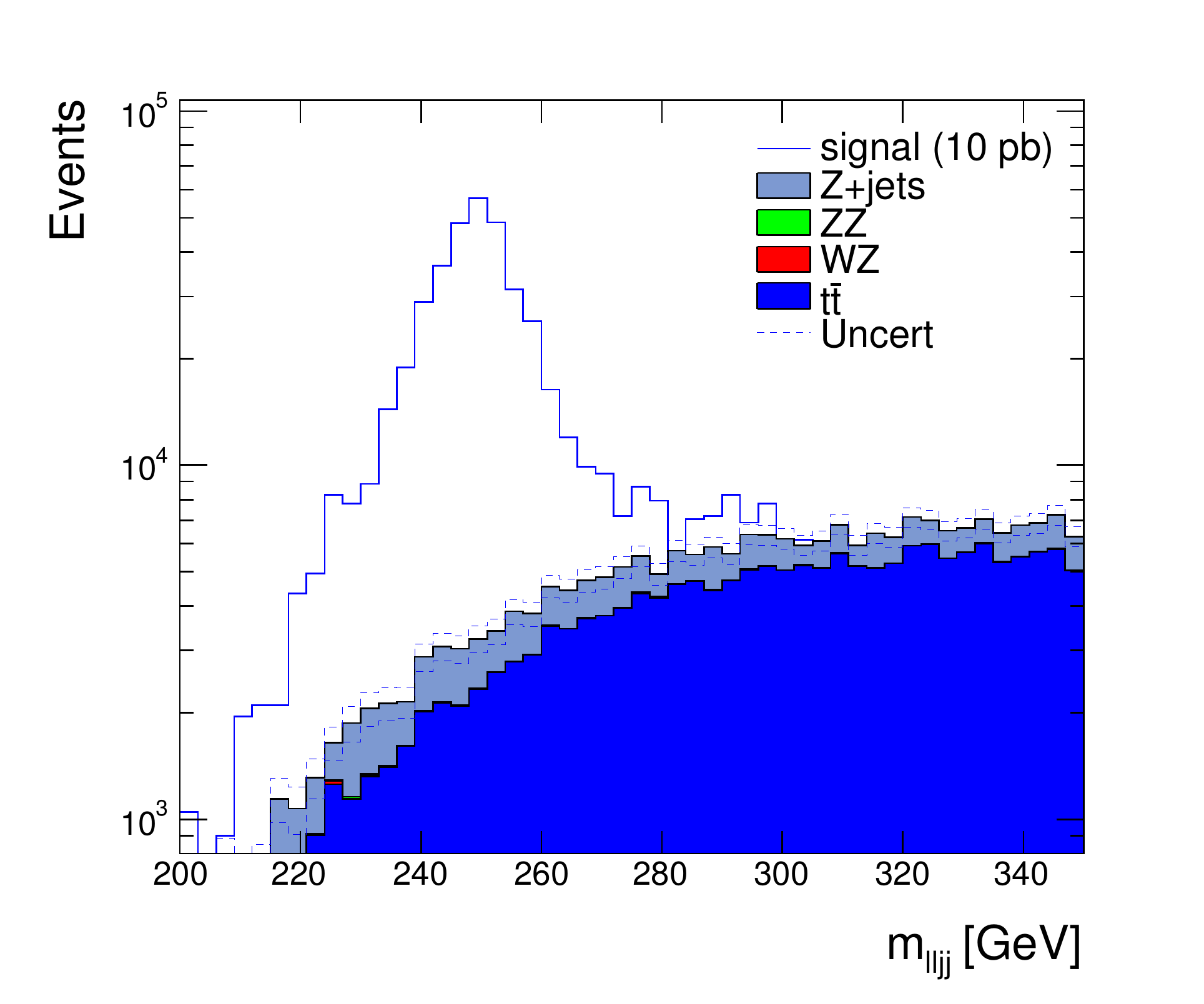}
\includegraphics[width=2in]{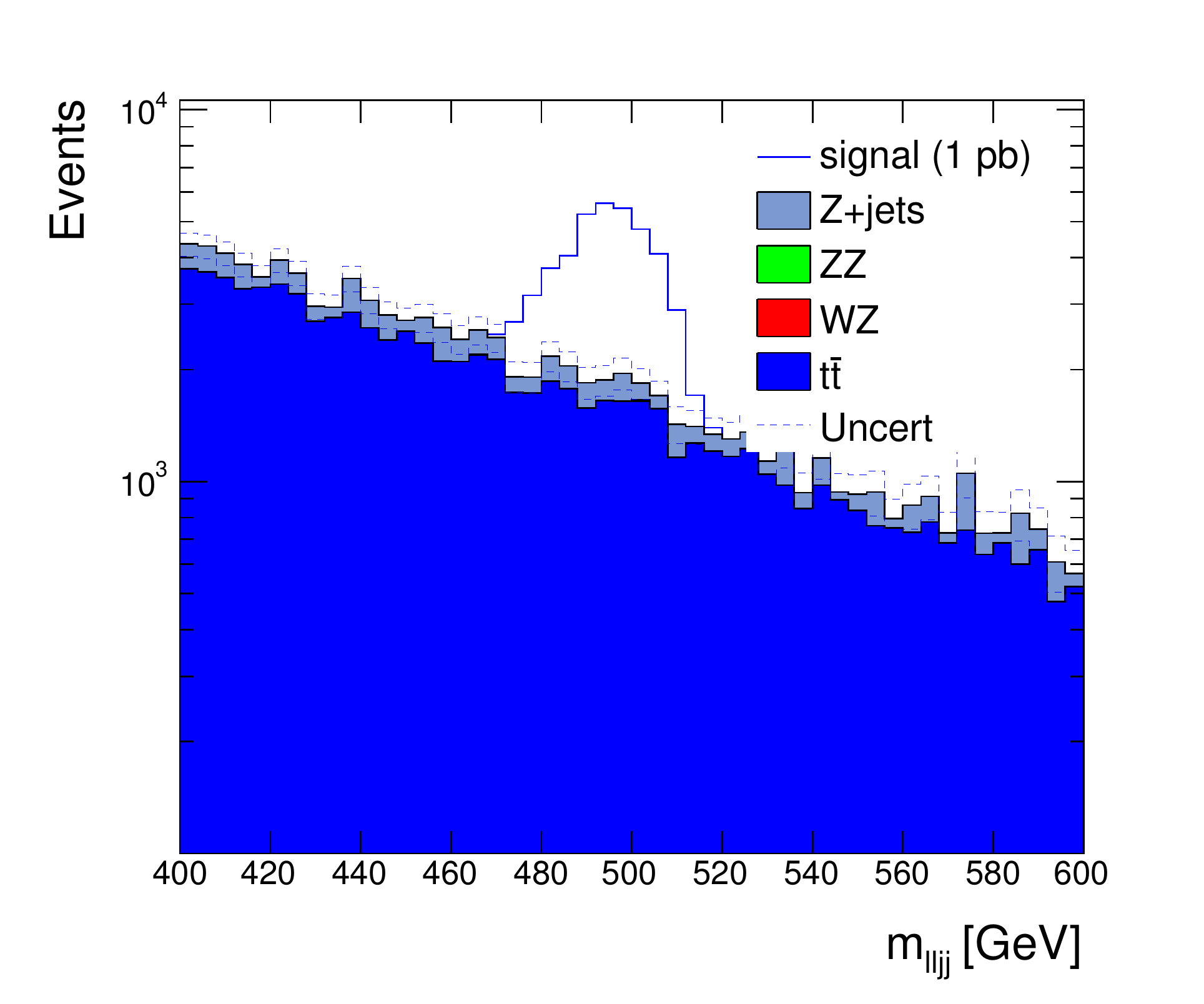}
\caption{In the $\ell(\ell jj)$ topology, the distribution of $m_{\ell\ell jj}$ in signal and background events for two example mass points, after requirements on $m_{\ell jj}$ and $m_{\ell\ell}$ in $pp$ collisions at $\sqrt{s}=14$~TeV with $\mathcal{L}=300$~fb$^{-1}$. Top shows the case of  $m_{Z'}=250$ GeV, $m_{L}=100$ GeV; bottom shows the case of  $m_{Z'}=500$ GeV, $m_{L}=200$ GeV.  
}\label{fig:ex2}
\end{figure}

\begin{figure}
\includegraphics[width=2in]{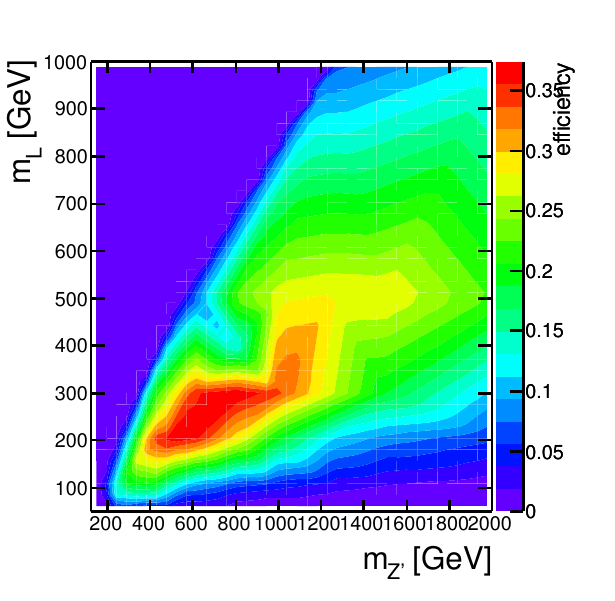}
\includegraphics[width=2in]{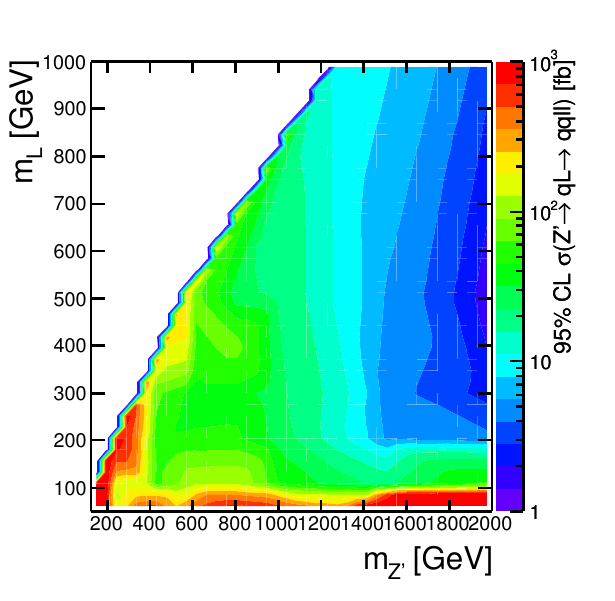}
\caption{In the $\ell(\ell jj)$ topology, selection efficiency and expected cross-section upper limits versus $m_{Z'}$ and  $m_L$ at $\sqrt{s}=14$~TeV with $\mathcal{L}=300$~fb$^{-1}$. For large $m_{Z'}-m_{L}$, the efficiency drops due to large transverse momentum of the $L$, which leads to small opening angles of the $L$ decay products.}
\label{fig:lim_l_ljj}
\end{figure}

In addition to the preselection requirements above, we select events
with reconstructed mass values close to the true values, $m_{\ell\ell j}
\in [m_{L}-100,m_{L}+50]$, and $m_{lljj}
\in [m_{Z'}-250,m_{Z'}+100]$. The requirement that $m_{\ell\ell}>120$ GeV
suppresses the $Z$-boson+jets background. Example distributions after
$m_{\ell jj}$ and $m_{\ell\ell}$ requirements can be seen for two examples in Fig~\ref{fig:ex2}.
 Efficiency of the final selection and expected upper limits on
 $\sigma(pp\rightarrow Z'\rightarrow \ell L \rightarrow
 \ell (\ell jj))$ can be seen in Fig~\ref{fig:lim_l_ljj}.

\begin{figure}
\includegraphics[width=2in]{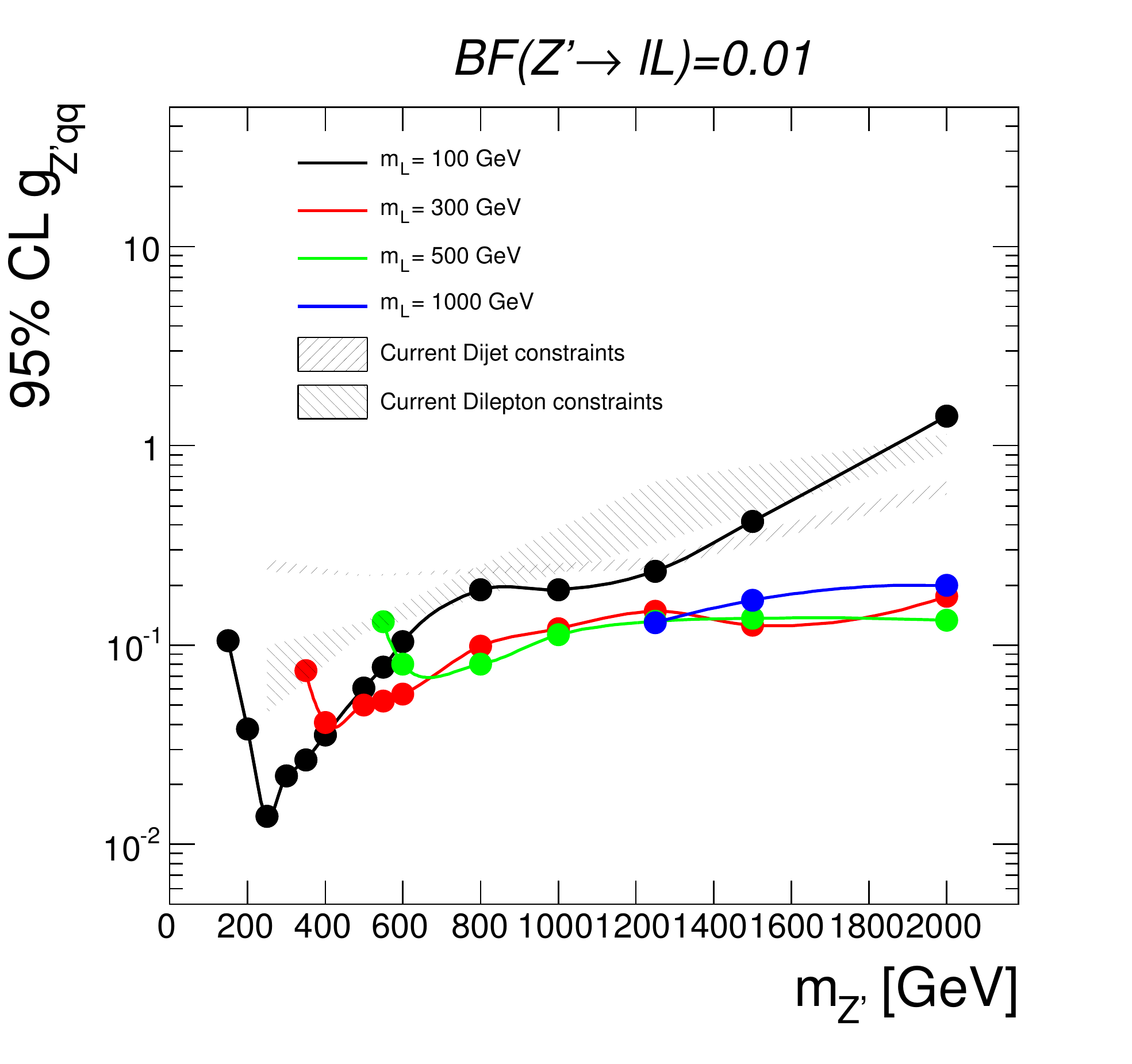}
\includegraphics[width=2in]{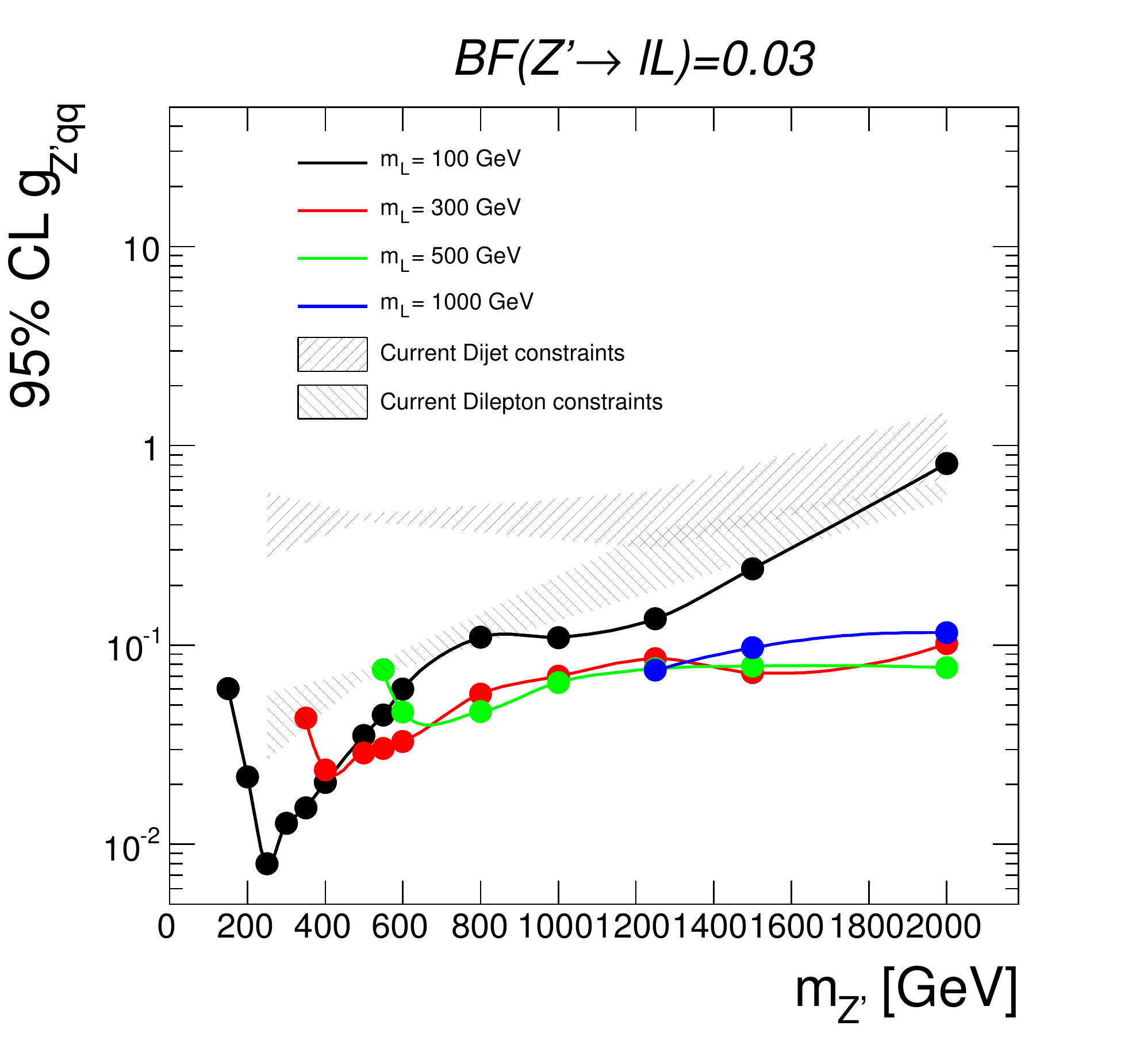}
\caption{ In the $\ell(\ell jj)$ topology, expected upper limits on the coupling $g_{Z'gg}$ versus $m_{Z'}$ and
  $m_L$ for two choices of $BF(Z'\rightarrow \ell L)$ at $\sqrt{s}=14$~TeV with
  $\mathcal{L}=300$~fb$^{-1}$.  The shaded region shows the
  current limits on the coupling from other topologies (see text)
  where the width of the band reflects the variation with assumed
  $m_{L}$.}
\label{fig:lim_l_ljj_g}
\end{figure}

As mentioned earlier, there are additional constraints coming from both dijet and dilepton decays of the $Z'$.  We use the ATLAS search for a dileptonic resonance, using 20 fb$^{-1}$ of data \cite{atlas2l}.  The relevant dijet resonance search \cite{cdf2j,cdf2jb,cms2j} depends upon the mass \cite{Dobrescu:2013cmh}.  All of these constraints are shown together in Fig~\ref{fig:lim_l_ljj_g}, over most of the parameter space considered the analysis outlined above provides the strongest constraint.

\subsection{$j(j\ell\ell)$ topology}

\begin{figure}
\includegraphics[width=2in]{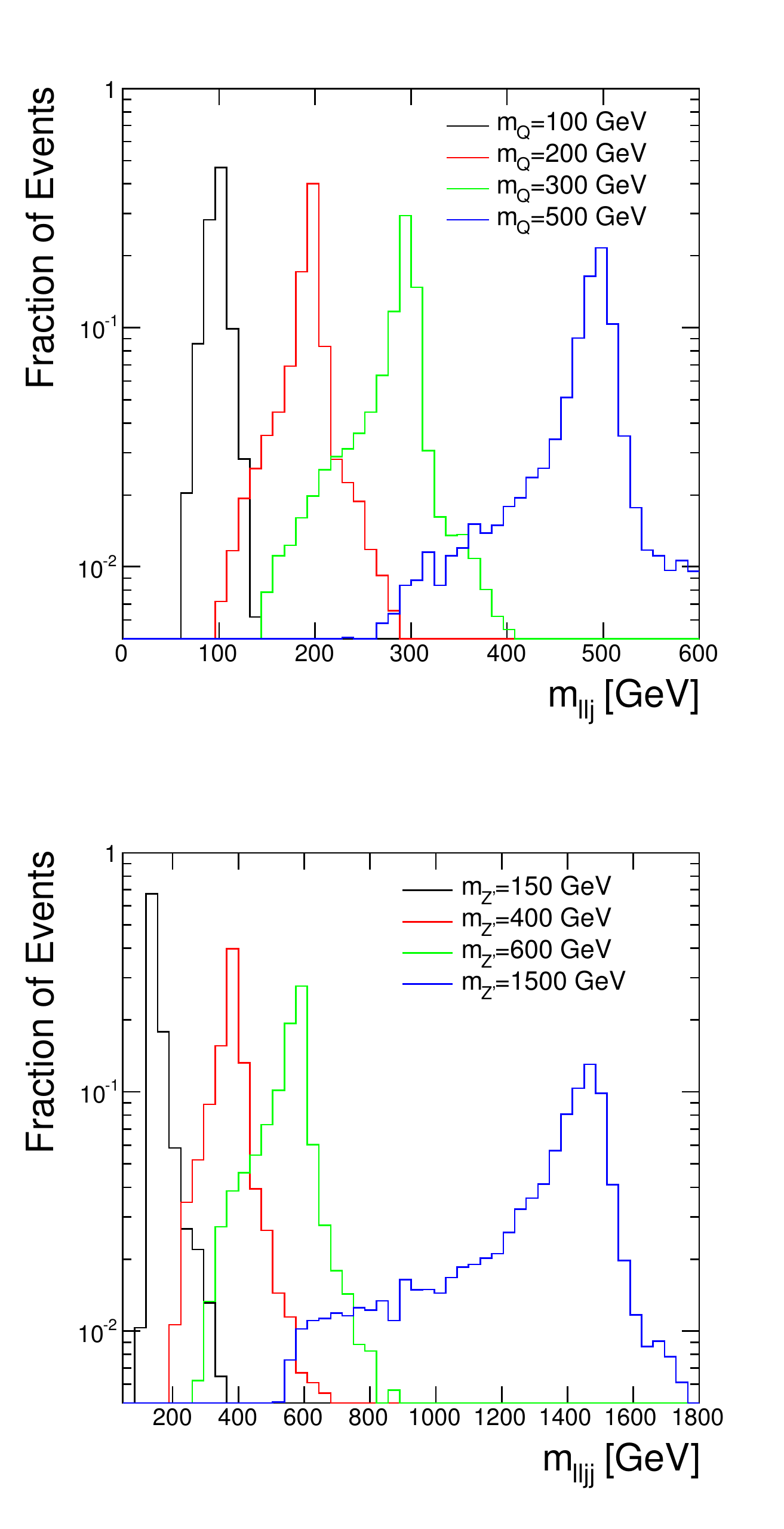}
\caption{In the $j(j\ell\ell)$ topology, distribution of reconstructed invariant $j\ell\ell$ and
  $\ell\ell jj$ masses for several values of  $m_{Z'}$
  and $m_{Q}$. Normalization is arbitrary.}
\label{fig:mass_j_jll}
\end{figure}

Resonance reconstruction in the $j(j\ell\ell)$ topology begins with
the identification of the two leptons produced in $Q\rightarrow j
\ell\ell$ decay, for which there are no ambiguities. The jet with smallest $\Delta R(j,\ell\ell)$ is chosen to
form $m_{Q}=m_{j\ell\ell}$, and the jet with largest  $\Delta
R(j,j\ell\ell)$ is chosen to form $m_{Z'}=m_{\ell\ell jj}$.  Examples of reconstructed masses are shown in Fig~\ref{fig:mass_j_jll}.

\begin{figure}
\includegraphics[width=2in]{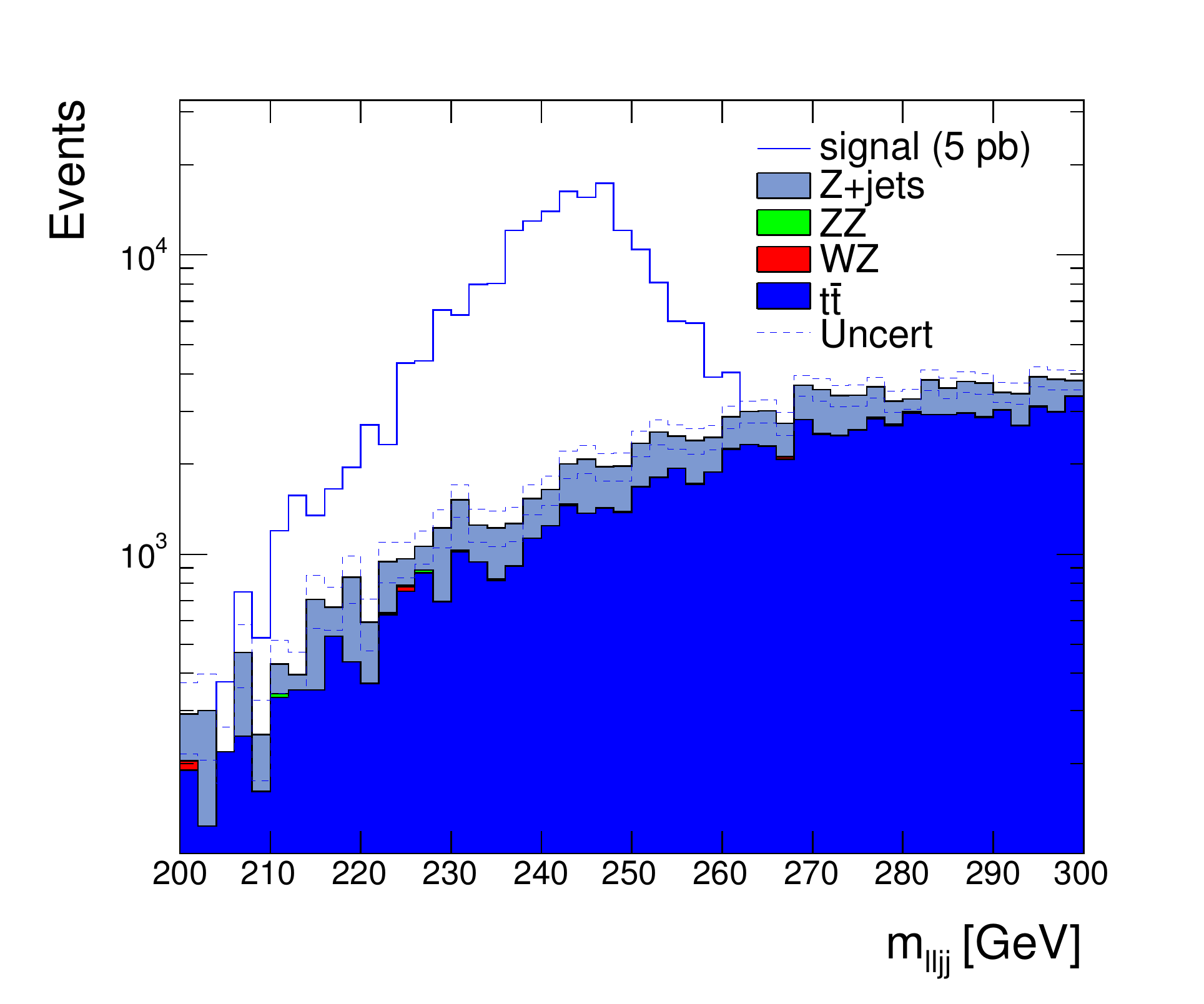}
\includegraphics[width=2in]{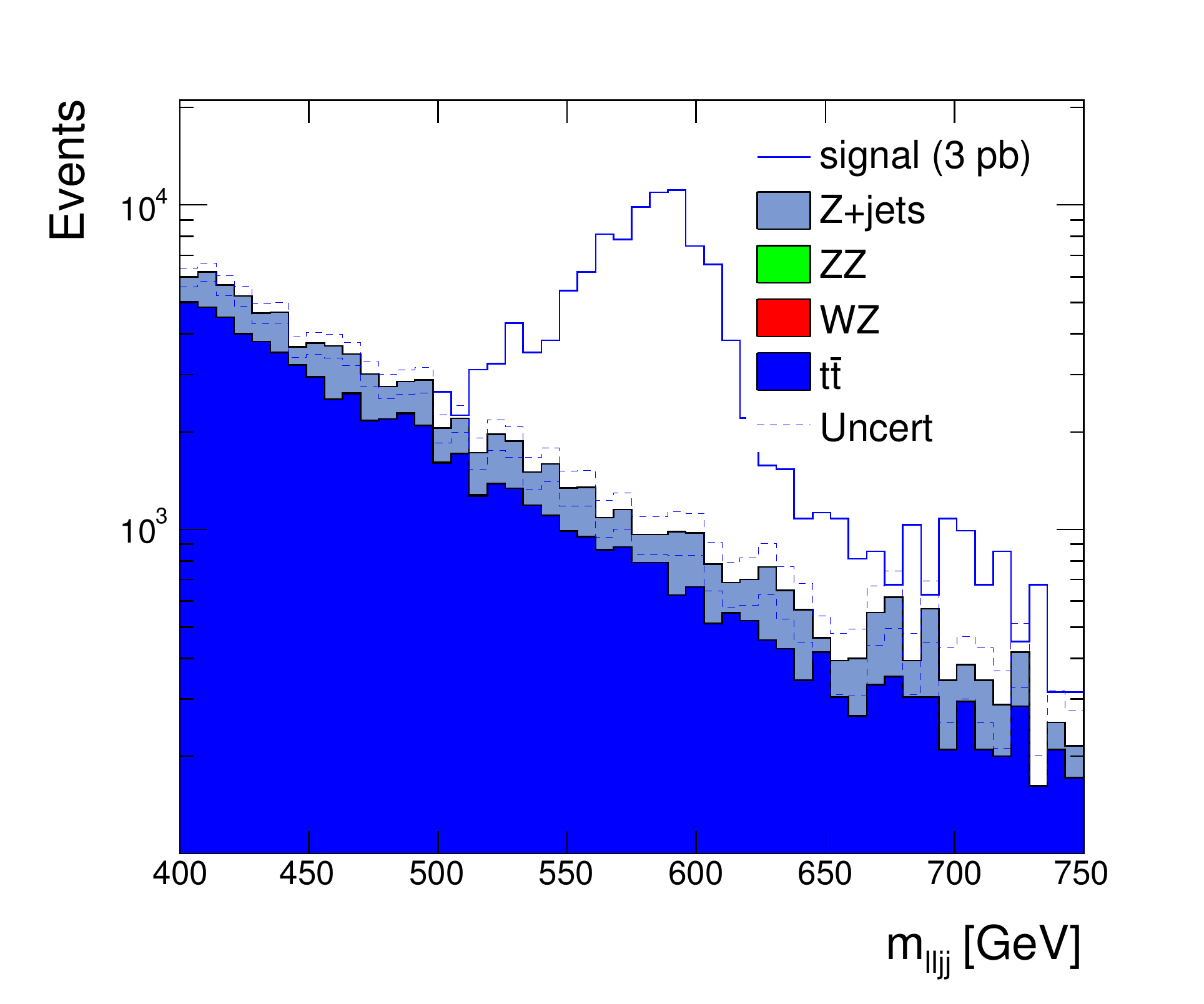}
\caption{In the $j(j\ell\ell)$ topology, distribution of $m_{\ell\ell jj}$ in signal and background
  events for two example mass points, after requirements on
  $m_{j\ell\ell}$ and $m_{\ell\ell}$ at $\sqrt{s}=14$~TeV with
  $\mathcal{L}=300$~fb$^{-1}$. Top shows the case of  $m_{Z'}=250$ GeV,
  $m_{Q}=200$ GeV; bottom shows the case of  $m_{Z'}=600$ GeV,
  $m_{Q}=200$ GeV.  In both cases, an arbitrary value of $\sigma(pp\rightarrow Z'\rightarrow qQ\rightarrow
 jj\ell\ell)$  is assumed.}
\label{fig:ex3}
\end{figure}

\begin{figure}[p]
\includegraphics[width=2in]{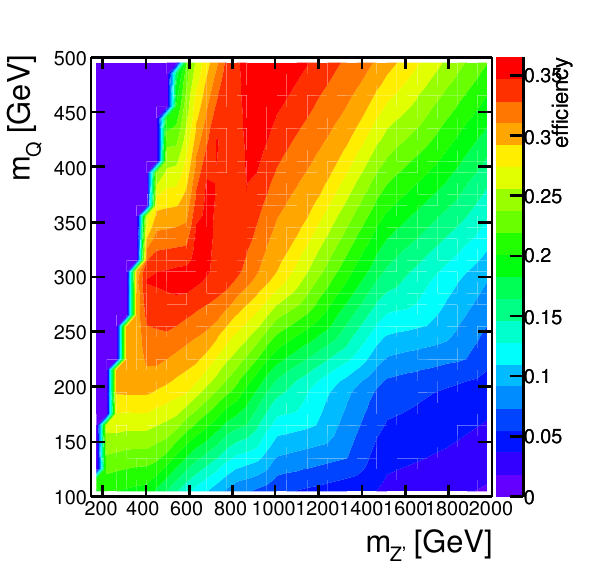}\\
\includegraphics[width=2in]{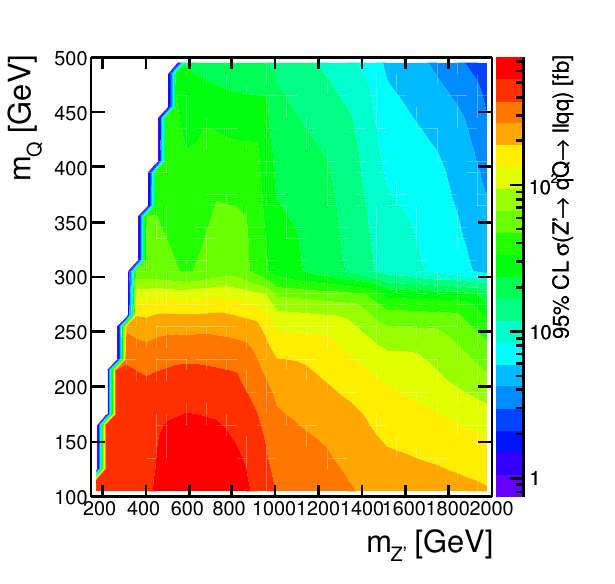}
\caption{In the $j(j\ell\ell)$ topology, selection efficiency (top) and expected cross-section upper limits (bottom) versus $m_{Z'}$ and
  $m_Q$ at $\sqrt{s}=14$~TeV with
  $\mathcal{L}=300$~fb$^{-1}$. For large $m_{Z'}-m_{Q}$, the efficiency drops due to large
  transverse momentum of the $Q$, which leads to small opening angles
  of the $Q$ decay products.}
\label{fig:lim_j_jll}
\end{figure}

In addition to the preselection requirements above, we select events
with reconstructed mass values close to the true values, $m_{\ell jj}
\in [m_{Q}-100,m_{Q}+50]$, and $m_{lljj}
\in [m_{Z'}-250,m_{Z'}+100]$.  Again, a requirement of $m_{\ell\ell}>120$ GeV
suppresses the $Z$-boson+jets background. Example distributions after
$m_{\ell\ell j}$ and $m_{\ell\ell}$ requirements can be seen for two examples in Fig~\ref{fig:ex3}.  The efficiency of the final selection and expected upper limits on
 $\sigma(pp\rightarrow Z'\rightarrow j Q \rightarrow
 j (j \ell \ell))$ can be seen in Fig~\ref{fig:lim_j_jll}.

\begin{figure}
\includegraphics[width=2in]{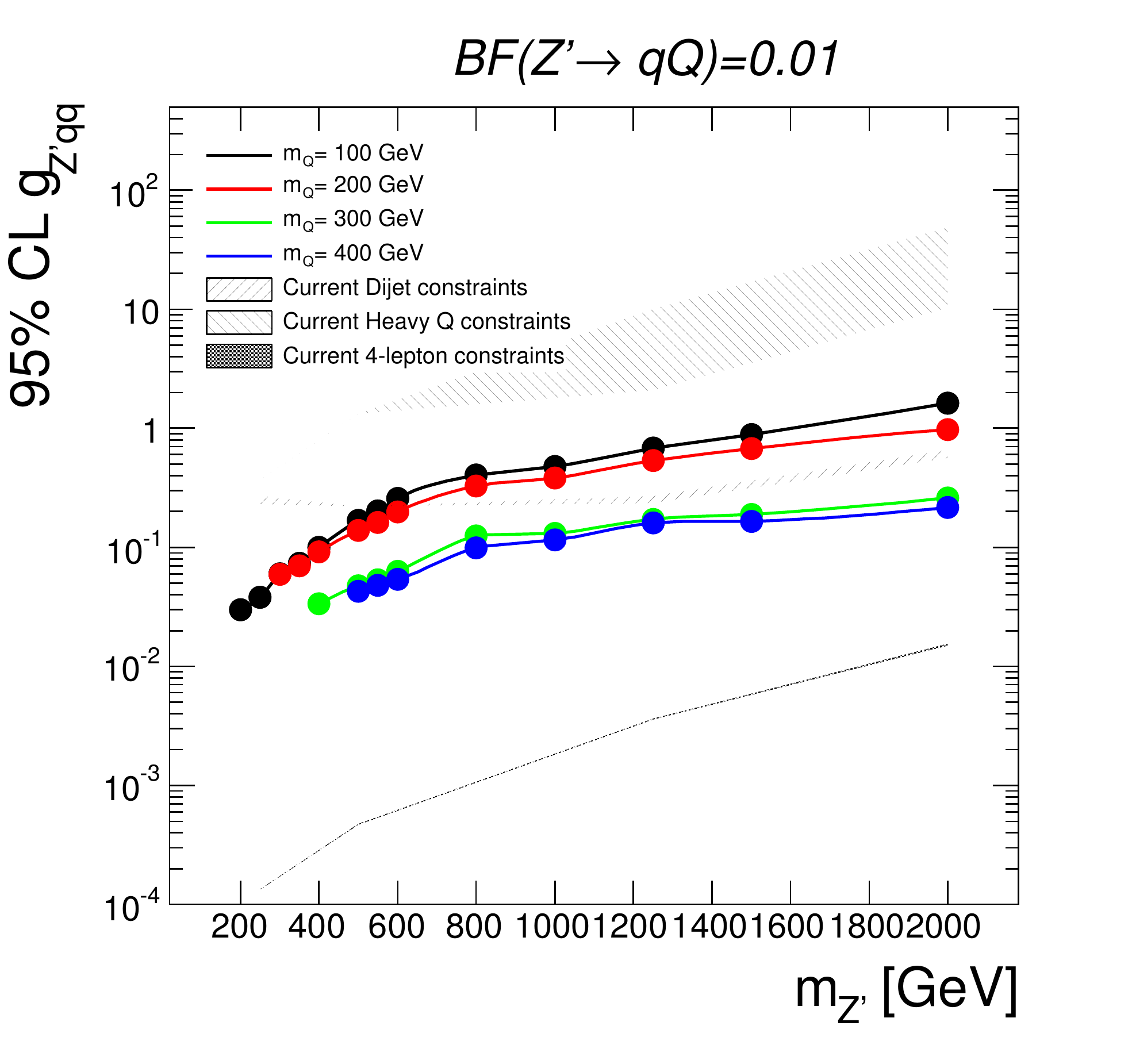}
\includegraphics[width=2in]{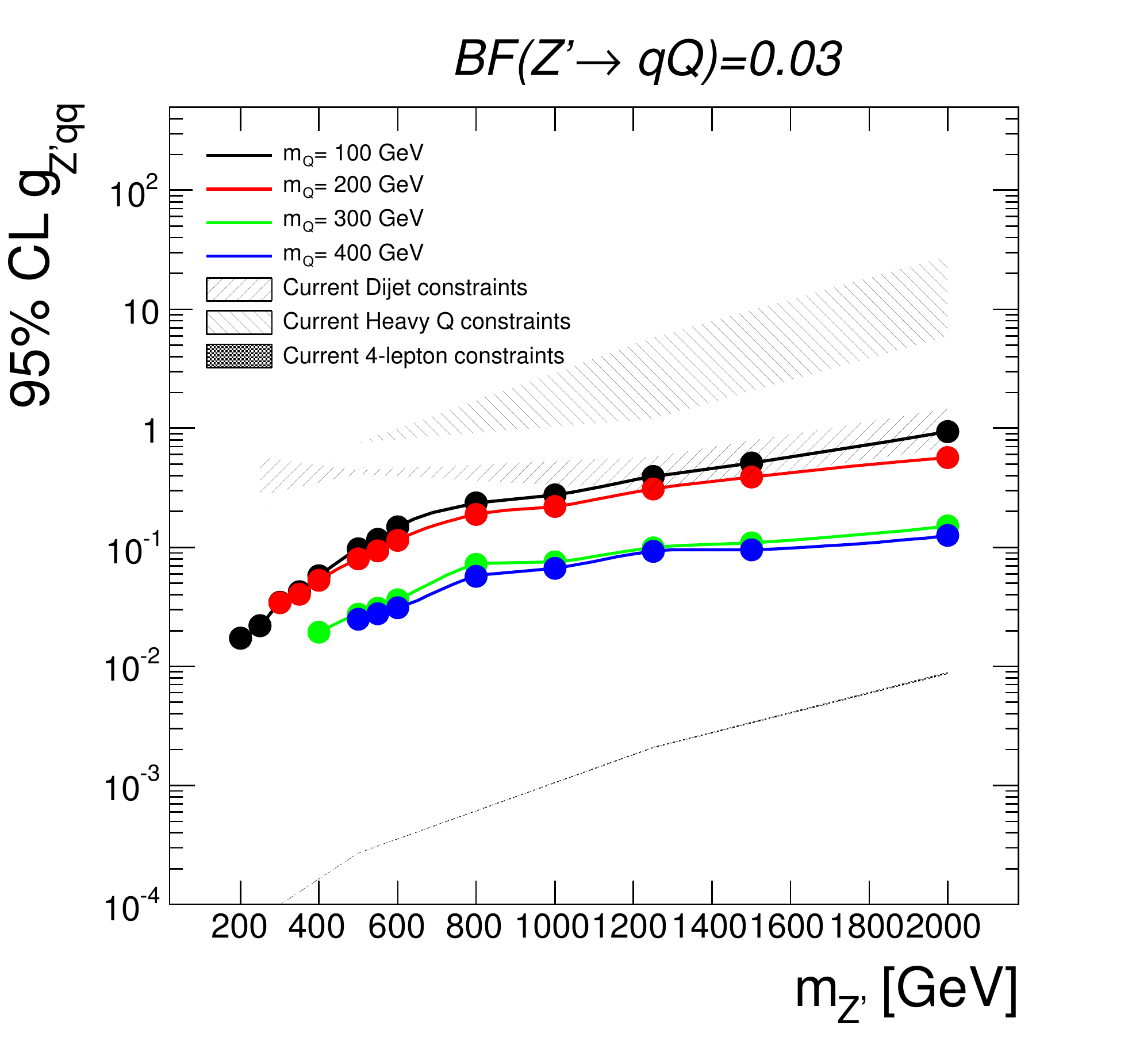}
\caption{Expected upper limits on the coupling $g_{Z'gg}$ versus $m_{Z'}$ and
  $m_Q$ for two choices of $BF(Z'\rightarrow qQ)$ at $\sqrt{s}=14$~TeV with
  $\mathcal{L}=300$~fb$^{-1}$.  The shaded region shows the
  current limits on the coupling from other topologies (see text)
  where the width of the band reflects the variation with assumed
  $m_{Q}$.}
\label{fig:lim_j_jll_g}
\end{figure}

As before there are other search modes for this model in related final states that place constraints on the same couplings.  There are constraints coming from dijet decays of the $Z'$ \cite{cdf2j,cdf2jb,cms2j}.  There is a constraint from an ATLAS search for a heavy quark~\cite{atlasQ}, using $1.04$ fb$^{-1}$ of 7 TeV data, where the singly produced heavy quark decays to a light quark and a leptonic $Z$.  Finally, there is a very strong constraint from LHC multilepton searches \cite{cmsmultilepton,atlas4l} on the pair production of $Q$ where each $Q$ decays to a jet and two leptons.  For the points in parameter space where the decay of $Z'\rightarrow Q\bar{Q}$ is kinematically accessible this is the strongest constraint, but if $M_Q>M_{Z'}/2$ the multilepton final state is suppressed by three body phase space and the mode searched for in this paper becomes an important constraint.  In Fig~\ref{fig:lim_j_jll_g} we show the limit on the coupling $g_{Z'qq}$ coming from $j(j\ell\ell)$ as well as the strongest, over all $M_Q$ at each $M_{Z'}$, of these other constraints.  Over all of the parameter space considered the heavy quark constraints are weakest, and the analysis described above is stronger than the dijet constraints for much, but not all, of the parameter space.  Although the multilepton constraint appears to be the strongest for every $Z'$ mass, it actually only applies for those parameter points where $M_Q<M_{Z'}/2$.

\section{Conclusions}

We have introduced a new approach, topological models, to systematically search for new
physics, which, in the absence of the discovery of a theoretically
predicted new particle, can point to new experimental search directions.  Like the simplified model approach, we advocate for minimal model descriptions to aid in the search for new phenomena.  However, rather than one or two simple models for a given final state, the topological models approach aims to cover the complete space of topologies for a particular final state.  By investigating all possible kinematic combinations of final state particles, whether they be motivated by existing models or not, additional discovery potential is unearthed.  In addition to the completeness of this approach, the characterization by final state resonance structure helps organize the presentation of experimental results.

As an example, we consider the final state of $\ell\ell jj$.  Some of the topologies that have been previously studied have only been analyzed under restrictive assumptions about the resonance masses, we generalize this.  Furthermore, we study those topologies, of the five possible, that have not been studied before.  We propose analysis techniques and study sensitivity for a 14 TeV LHC with $\mathcal{L}=300$ fb $^{-1}$.  Both this generalization of existing searches and the addition of new topologies, fill in gaps in experimental analyses where there exists potential for discovery.  Repeating this procedure on actual LHC data and in more final states can potentially lead to untapped discoveries.

\subsection{Acknowledgements}

We acknowledge useful conversations with Tim Tait, Roni Harnik and
Bogdan Dobrescu. We are grateful to Felix Yu and Flip Tanedo for insightful commentary.
DW is supported by grants from the Department of Energy
Office of Science and by the Alfred P. Sloan Foundation.  Fermilab is operated by Fermi Research Alliance, LLC under Contract No. DE-AC02-07CH11359 with the United States Department of Energy.




\clearpage

\end{document}